\documentclass[fleqn,usenatbib]{mnras}

\usepackage{newtxtext,newtxmath}

\usepackage[T1]{fontenc}

\usepackage{graphicx}
\usepackage{subcaption}

\DeclareRobustCommand{\VAN}[3]{#2}
\let\VANthebibliography\thebibliography
\def\thebibliography{\DeclareRobustCommand{\VAN}[3]{##3}\VANthebibliography}

\usepackage{graphicx}	
\usepackage{amsmath}	

\newcommand\teff{T$_{\rm eff}$~}
\newcommand\feh{$[$Fe/H$]$~}

\title[H$\alpha$ Age-Activity Relation for Solar-Type Stars]{Fine Structure of the Age-Chromospheric Activity Relation in Solar-Type Stars: II. H$\alpha$ Line\thanks{Based on observations collected at Observat\'orio do Pico dos Dias (OPD), operated by the Laborat\'orio Nacional de Astrof\'s{\i}ca, MCTI, Brazil.}}

\author[P. V. Souza dos Santos et al.]{
P. V. Souza dos Santos$^{1}$\thanks{E-mail: paulovss.astro@gmail.com},
G. F. Porto de Mello$^{1}$,
E. Costa-Bhering$^{1}$,
D. Lorenzo-Oliveira$^{2}$,
\newauthor{F. Almeida-Fernandes$^{3}$,
L. Dutra-Ferreira$^{4}$,
and I. Ribas$^{5,6}$}
\\
$^{1}$Universidade Federal do Rio de Janeiro, Observatório do Valongo, Ladeira do Pedro Antonio 43, CEP: 20080-090 Rio de Janeiro, RJ, Brazil
\\
$^{2}$Laborat\'orio Nacional de Astrof\'{\i}sica, Rua dos Estados Unidos, 154, Itajubá, MG, Brazil, 37504-364\\
$^{3}$Departamento de Astronomia, Instituto de Astronomia, Geofísica e Ciências Atmosféricas da USP, Cidade Universitária, 05508-900, São Paulo, SP, Brazil\\
$^{4}$CAp-UERJ, R. Bar\~ao de Itapagipe, 96, Rio de Janeiro, RJ, Brazil, 20261-005\\
$^{5}$Institut de Ciències de l’Espai (ICE, CSIC), Campus UAB, c/ Can Magrans s/n, 08193 Bellaterra, Barcelona, Spain\\
$^{6}$Institut d’Estudis Espacials de Catalunya (IEEC), c/ Gran Capità 2–4, 08034 Barcelona, Spain
}

\date{Accepted XXX. Received YYY; in original form ZZZ}

\pubyear{2024}

\begin{document}


\label{firstpage}
\pagerange{\pageref{firstpage}--\pageref{lastpage}}
\maketitle

\begin{abstract}
Excess chromospheric emissions within deep photospheric lines are effective proxies of stellar magnetism for FGK stars. This emission decays with stellar age and is a potential determinant of this important stellar quantity. We report absolutely calibrated H$\alpha$ chromospheric fluxes for 511 solar-type stars in a wide interval of precisely determined masses, $[$Fe/H$]$, ages, and evolution states from high S/N, moderately high$-$resolution spectra. The comparison of H$\alpha$ and H+K chromospheric fluxes reveals a metallicity bias (absent from H$\alpha$) affecting \ion{Ca}{ii} H+K fluxes thereby metal-rich stars with deep line profiles mimic low chromospheric flux levels, and vice versa for metal-poor stars. This bias blurs the age-activity relation, precluding age determinations for old, inactive stars unless mass and $[$Fe/H$]$ are calibrated into the relation. The H+K lines being the most widely studied tool to quantify magnetic activity in FGK stars, care should be exercised in its use whenever wide ranges of mass and $[$Fe/H$]$ are involved. The H$\alpha$ age-activity-mass-metallicity calibration appears to be in line with the theoretical expectation that (other parameters being equal) more massive stars possess narrower convective zones and are less active than less massive stars, while more metal-rich stars have deeper convective zones and appear more active than metal-poorer stars. If regarded statistically in tandem with other age diagnostics, H$\alpha$ chromospheric fluxes may be suitable to constrain ages for FGK stars with acceptable precision.
\end{abstract}




\section{Introduction}

Age is among the most difficult stellar parameters to gauge with any confidence, and only for the Sun can it be determined in a hypothesis-free approach from primordial meteorite differentiates \citep{gancarz1977initial}. Reliable age determinations are essential ingredients to understanding the chemo-dynamical evolution of the Galaxy and other stellar systems. A problem that has recently come to the front of astrophysical research is the necessity to determine the ages of star systems harboring potentially habitable exoplanets, both to understand their probable atmospheric evolution and to provide context for the interpretation of biosignatures \citep{holland2006oxygenation,gialluca2021characterizing,mendez2021habitability}. The evolution of non-thermal, chromospheric excess fluxes (usually referred to as chromospheric activity) in the core of deep photospheric lines is widely regarded as well-connected to the aging of low-mass stars and the evolution of stellar magnetism. 


Attempts to characterize the chromospheric flux as a function of stellar age began with the seminal work of \cite{skumanich1972time}, using the H+K lines of \ion{Ca}{ii} of a few stars of open clusters and kinematic groups. The decades-long Mount Wilson project \citep{baliunas1995chromospheric} has been monitoring the \ion{Ca}{ii} H+K $<$S$>$ index, which can be converted into the \text{log}(R'$_{\rm HK}$) index, defined as the absolute line excess flux (total line flux $-$ photospheric flux) normalized to the bolometric flux \citep{linsky1979stellar,noyes1984rotation}. The \text{log}(R'$_{\rm HK}$) is well-established as the standard metric to estimate stellar ages through age-chromospheric activity relations \citep{mamajek2008improved}. \cite{soderblom1985survey}, \cite{barry1987chromospheric}, \cite{barry1988chromospheric} and \cite{soderblom1991chromospheric} followed on the use of these lines as chromospheric indicators of magnetic activity and age determinants. 

More recently, \cite{pace2004age} and \cite{pace2013chromospheric} argued that the H+K age-activity relation is discernible only for stars younger than about $\sim$2 Gyr. \cite{lorenzo2016age} disputed this result and built a more comprehensive H+K age-activity relation including stellar mass and metallicity as calibrating terms and were able to recover precise stellar ages up to $\sim$6 Gyr. Mass and metallicity are stellar structural parameters expected to dictate the extension of the convective zone and, consequently, the power of the dynamo effect, directly influencing the timescale of magnetic evolution. For example, by using exclusively solar twins, thus narrowing the range of stellar mass and metallicity, \cite{lorenzo2018solar} showed the age-activity relation to remain sensitive to stellar ages up to $\sim$7 Gyr. The dependence of chromospheric activity evolution with stellar metallicity was also theoretically verified by \cite{amard2020impact}, reporting that metal-rich stars slow their rotation more efficiently than metal-poor stars for ages larger than 1 Gyr.

Other strong photospheric lines, such as H$\alpha$ and the infrared triplet lines of \ion{Ca}{ii}, are also useful chromospheric activity diagnostics. H$\alpha$ has a lower chromospheric contrast than the H+K transitions \citep{pasquini1991h}, which makes the chromospheric excess flux harder to discern. Unlike the H+K lines, for which extensive surveys exist \citep[e.g.][]{daSilva2021stellar}, H$\alpha$ has received far less attention as a magnetic activity and chromospheric emission diagnostic, even though spectroscopic surveys covering this spectral range, such as GALAH \citep{buder2018galah}, have recently become available. 

Despite their lower chromospheric contrast, H$\alpha$ fluxes have the advantage of being much less influenced by stellar magnetic cycles and transient phenomena such as flares and starspots \citep[][hereafter, LPM05]{lyra2005fine}, which can show variations in \text{log}(R'$_{\rm HK}$) even for very inactive stars \citep{daSilva2021stellar}. The H$\alpha$ line profile, in addition, is nearly insensitive to stellar metallicity \citep{fuhrmann1993balmer}, a parameter expected to directly affect the H+K chromospheric flux measurements. This metallicity dependence is also expected for other chromospheric diagnostics based on metal lines, such as the \ion{Ca}{ii} infrared triplet \citep[also far less studied than the H+K lines:][]{chmielewski2000infrared,busa2007ii,lorenzo2016fine} and the \ion{Mg}{ii} lines.

Attempts to employ H$\alpha$ as a chromospheric indicator of age are very scarce in the literature. \cite{herbig1985chromospheric} carried out a pioneer study for a sample of 40 stars. LPM05 built a multiparametric age-activity relation from H$\alpha$ chromospheric fluxes for 175 stars, including mass and metallicity as regressive variables. These authors were the first to confirm observationally theoretical predictions connecting convective properties \citep[as described by the Rossby number:][]{barnes2010angular,noyes1984rotation} to stellar mass, metallicity, and the timescale of chromospheric activity evolution. The Rossby number connects rotational evolution to the extension of the convective zone: LPM05 highlighted that stellar structural parameters that drive convective efficiency should be explicitly taken into account in the derivation of age-activity relations, an assertion later reinforced by \cite{lorenzo2016age}.

More recently, \cite{douglas2014factory} studied the H$\alpha$ chromospheric emission, quantified as line equivalent widths for Hyades and Praesepe stars, reporting a dependence with the Rossby number. \cite{sissa2016halpha} investigated the influence of H$\alpha$ chromospheric fluxes on the measurement of ultraprecise radial velocities and built an age-activity relation which, however, they found to be insensitive to stellar age beyond 1.5 Gyr. Finally, \cite{zhang2019stellar} used the LAMOST survey to identify open clusters and study chromospheric emission in both the \ion{Ca}{ii} K line and H$\alpha$ for $\sim$1,100 stars. They report that their age-activity relations estimate stellar ages within 40\% accuracy for log (R'$_{\rm HK}$) and 60\% using \text{log}(R'$_{\rm H\alpha}$). 

Towards establishing the usefulness of other spectroscopic indicators of chromospheric activity, besides H$+$K, for solar-type stars, \cite{lorenzo2016fine} (hereafter, Paper I) reported absolute chromospheric fluxes for the \ion{Ca}{ii} triplet lines calibrated absolutely by model atmosphere fluxes. In the present work, we expand on the work of LPM05, enlarging the stellar sample and deriving total and chromospheric absolute fluxes in physical units (erg $\text{cm}^{-2} \text{ s}^{-1}$), through model atmosphere theoretical absolute fluxes. We analyzed the H$\alpha$ chromospheric fluxes in terms of their correlation to \ion{Ca}{ii} H+K chromospheric losses and their evolution in time.

This paper is organized as follows: we present the sample and data details in Section \ref{sec2}. In Sect. \ref{sec3} we discuss the determination of the stellar parameters: effective temperature (hereafter T$_{\rm eff}$), metallicity (hereafter $[$Fe/H$]$), surface gravity, mass, radius, luminosity, absolute magnitude and age. Details of the derivation of the total absolute flux in the H$\alpha$ line, as well as the subtraction of the photospheric component, are presented in Sect. \ref{sec4}. In Sect. \ref{sec5}, we quantify the biases affecting the chromospheric fluxes of the H+K and triplet lines and calibrate the H$\alpha$ chromospheric fluxes against stellar ages, for the subsample of stars with precise isochronal ages, followed by a summary of our conclusions in Sect. \ref{sec6}.

\section{Sample, Observations and Reduction}
\label{sec2}

Our data consists of spectra of 511 stars of F, G, and K spectral types, main sequence dwarfs and subgiants, and the vast majority of stars part of the solar neighbourhood (distances less than 50 parsecs to the Sun). Some sample objects possess particular relevance:
\begin{enumerate}
    \item members of young stellar associations (Tucana-Horologium, Beta-Pictoris) and open clusters (Hyades, Pleiades); their coeval nature implies well-constrained metallicity and age;
    \item stars previously known as chromospheric inactive stars \citep{henry1996survey,wright2004chromospheric}, tasked with setting the lowest absolute flux in H$\alpha$ at a given T$_{\rm eff}$ and defining the envelope of photospheric correction, as will be shown in Sect. \ref{sec4};
    \item metal-poor stars and cool stars (K dwarfs), objects that were underrepresented in our previous work (LPM05).
    \item stars for which asteroseismological ages are available as a fundamental check on the isochronal and chromospheric ages.
\end{enumerate} 

\begin{figure}
\centering
	\includegraphics[scale=0.26]{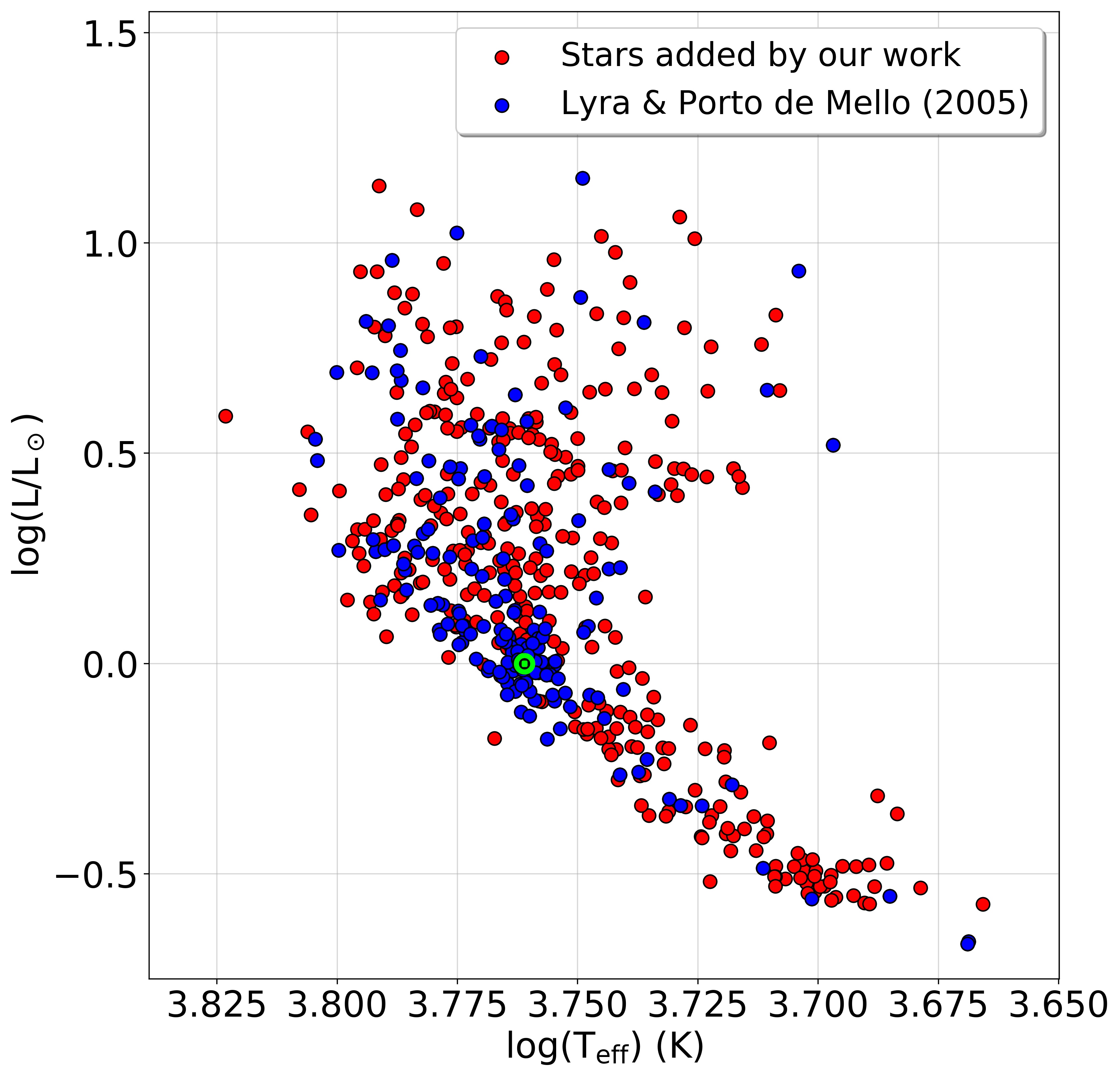}
    \caption{HR diagram comparing our sample with the sample of LPM05. The Sun is represented by the usual symbol, in green.}
    \label{fig:hrdiagram_lyra}
\end{figure}

\begin{figure}
\centering
\includegraphics[width=1\linewidth]{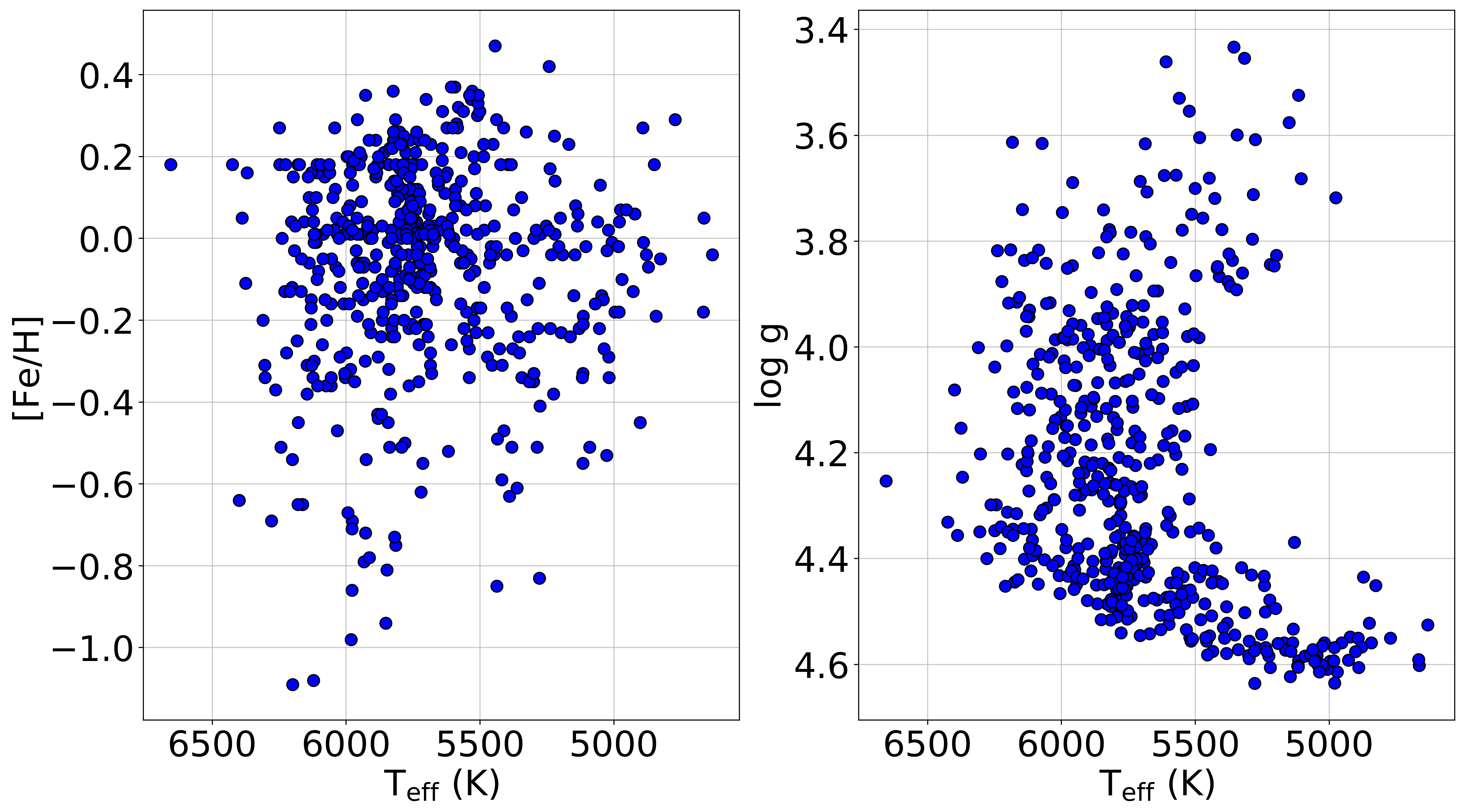}
    \caption{Distribution of T$_{\rm eff}$, $\log{g}$ and metallicity $[$Fe/H$]$ for the sample stars.}
    \label{fig:atmospheric_parameters}
\end{figure}

We show in Fig. \ref{fig:hrdiagram_lyra} the range of astrophysical parameters of the sample in a HR diagram, comparing with the sample of LPM05. Sample stars fill the range $4600 \lesssim \text{T}_{\text{eff}} \lesssim 6600$ K, $-1.1 \lesssim \text{[Fe/H]} \lesssim 0.5$ dex, $3.4 \lesssim \text{$\log{g}$} \lesssim 4.6$ dex, $0.2 \lesssim \text{L/L$_\odot$} \lesssim 14$. Stellar masses lies in the range $0.7 \lesssim \text{M/M$_\odot$} \lesssim 1.6$, centered around the solar value. Stellar ages span from a few Myr (the young stellar associations) up to more than 10 Gyr. The distribution of the atmospheric parameters can be seen in Fig. \ref{fig:atmospheric_parameters}. The final adopted values for all parameters shown in Fig. \ref{fig:hrdiagram_lyra} and \ref{fig:atmospheric_parameters} are described in detail in Sec. \ref{sec3}.

All spectra were obtained in the Observatório do Pico dos Dias (Laboratório Nacional de Astrofísica), using the 1.6 m Perkin-Elmer telescope and the coudé spectrograph. A total of 511 stars were observed, 235 of them at moderately high resolution ($R$ $\sim 30\,000$), 214 at moderate resolution ($R \sim 11\,000$), and 62 at both resolutions. In order to include solar flux spectra in our sample, we have observed galilean satellites and the Moon as solar proxies in many runs, exposed to very high S/N, the latter with its image stopped orthogonally to the slit width to a size comparable to the seeing disks of the stars.

We used IRAF\footnote{IRAF is distributed by the National Optical Astronomy
Observatories, which is operated by the Association of Universities for
Research in Astronomy, Inc. (AURA) under a cooperative agreement with the National
Science Foundation} to reduce the spectra, following the standard procedure. The mean value of S/N for the spectra is $\sim 170$: very few spectra have S/N $<$ 100, and the bulk of the data lie in the 100 $<$ S/N $<$ 300 range.

\section{Determination of Stellar Parameters}
\label{sec3}

The accurate determination of absolute chromospheric fluxes of the stars and their evolution with age calls for good-quality atmospheric parameters. We first compiled the atmospheric parameters (T$_{\text{eff}}$, $\log{g}$ and [Fe/H]) from recent literature. These works mostly derive these parameters spectroscopically, from the excitation and ionization equilibrium of Fe I and Fe II lines, and employ very heterogeneous model atmospheres, techniques and line lists. The full list of references utilized in the atmospheric parameters compilation is available in Table \ref{tab:atmospheric_parameters}. We strived to homogenize the data as described below, which also affects the scale of the \feh, mass, surface gravity, and age values.

\nocite{adibekyan2012chemical,aguilera2018lithium,bensby2014exploring,boesgaard1990chemical,boesgaard2004correlation,carrera2019open,chaffee1971abundances,da2012accurate,datson2015spectroscopic,dopcke2019ursa,favata1997fe,buder2019galah,gehren2004abundances}
\nocite{ghezzi2005,gratton2003abundances,gray2006contributions,kounkel2019close,luck2017abundances,luck2018abundances,mishenina2012activity,mishenina2013abundances,montes2018calibrating,paulson2003searching,paulson2006differential,del2005agei,del2005ageiii}
\nocite{de2008alpha,de2014photometric,ramirez2012lithium,ramirez2013oxygen,randich1999lithium,santos2004spectroscopic,santos2005spectroscopic,sitnova2015systematic,soto2018spectroscopic,tagliaferri1994photometric,trevisan2011analysis,valenti2005spectroscopic}
\nocite{wittenmyer2016pan,sousa2006spectroscopic}

\subsection{Photometric effective temperature}

\cite{giribaldi2019accurate} compared T$_{\rm eff}$ scales from photometry, the excitation \& ionization equilibria of Fe, and the fitting of models to H$\alpha$ profiles and concluded that the photometric IRFM \teff scale is fully compatible with the fundamental interferometric one \citep{casagrande2014towards}. We thus set our \teff scale according to the calibrations of \cite{casagrande2010absolutely} and \cite{casagrande2021galah}, using the colours:
\begin{enumerate}
    \item $B-V$ and $B_T-V_T$, retrieved from the HIPPARCOS catalogue \citep{van2007validation};
    
    \item $b-y$, retrieved from, primarily, \cite{olsen1983four},  \cite{olsen1993stromgren} and \cite{olsen1994large} (colours were transformed to the system of Olsen 1993, according to this author's prescriptions);
    
    \item $G-RP$, $BP-RP$ and $G-BP$, from Gaia DR2 \citep{2018A&A...616A...1G}.
\end{enumerate}
  
 Each of these six colours provides a \teff estimate, using also the \feh and $\log{g}$ figures taken straight from the literature. We adopted a weighted mean value of the effective temperatures ($\bar{T}$) using the calibrations errors (spanning from 62 to 93 K). The resulting uncertainty of each mean ($\sigma_{\bar{T}}$) was calculated using the standard deviation weighted by the calibration errors.
A 2$\sigma$ clipping procedure removed \teff values that clearly behaved as outliers, after which the mean \teff internal uncertainty value converged to 38 K.

\begin{figure}
  \begin{subfigure}[b]{0.49\linewidth}
    \centering
    \includegraphics[width=1\linewidth]{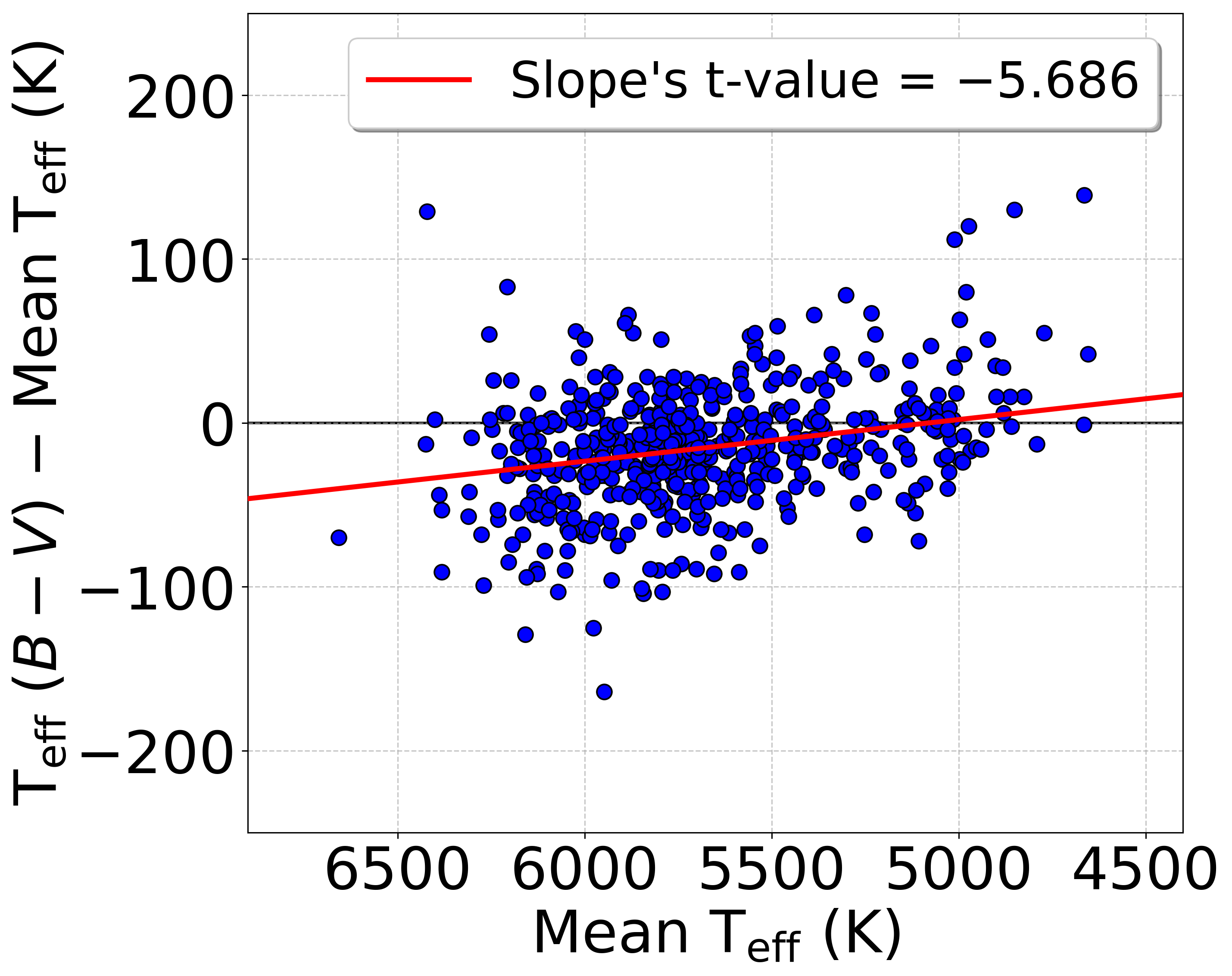} 
    \caption{}
    \vspace{2ex}
  \end{subfigure}
  \hspace{1ex}
  \begin{subfigure}[b]{0.49\linewidth}
    \centering
    \includegraphics[width=1\linewidth]{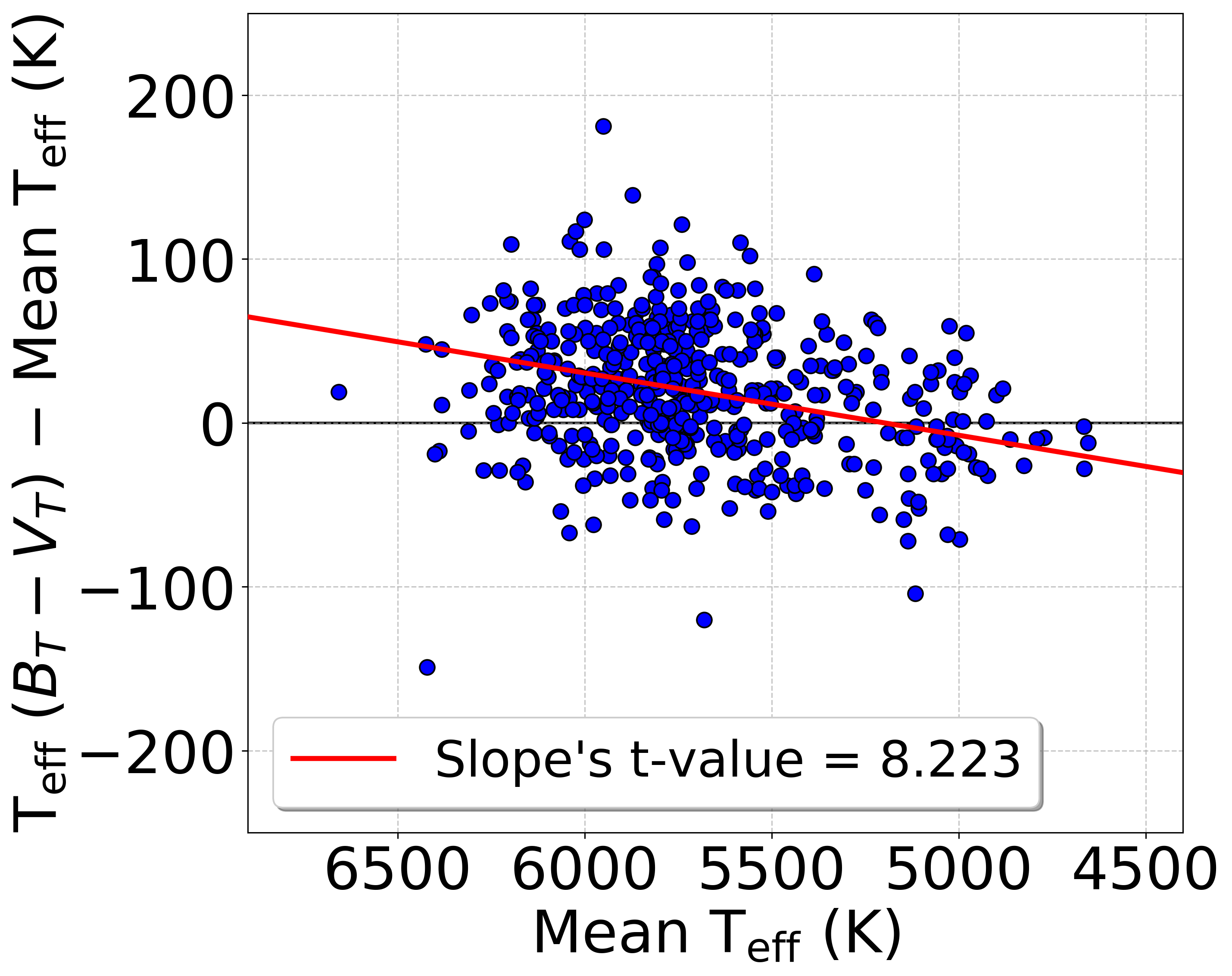} 
    \caption{}
    \vspace{2ex}
  \end{subfigure} 
  \begin{subfigure}[b]{0.49\linewidth}
    \centering
    \includegraphics[width=1\linewidth]{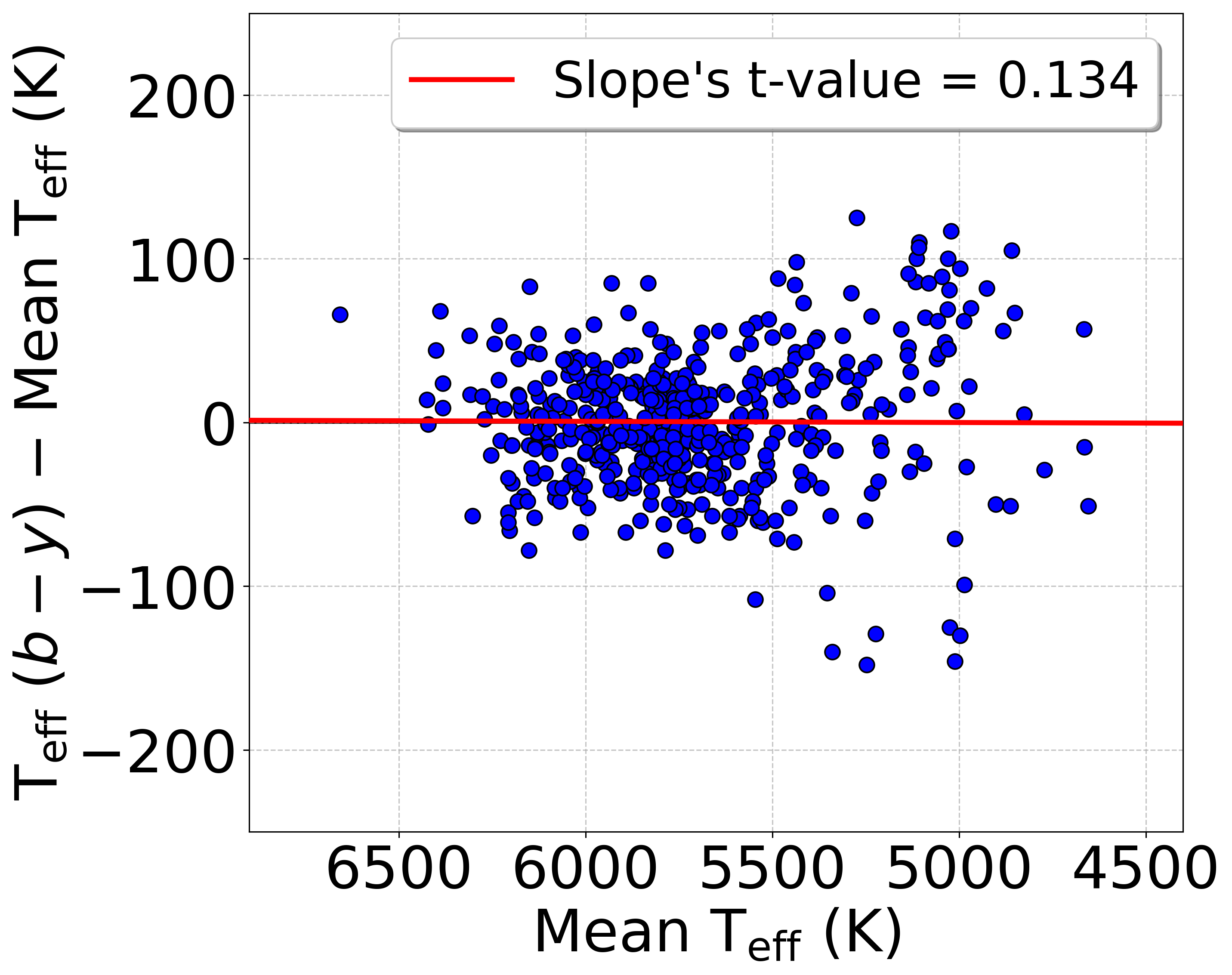} 
    \caption{}
    \vspace{2ex}
  \end{subfigure}
  \hspace{1ex}
  \begin{subfigure}[b]{0.49\linewidth}
    \centering
    \includegraphics[width=1\linewidth]{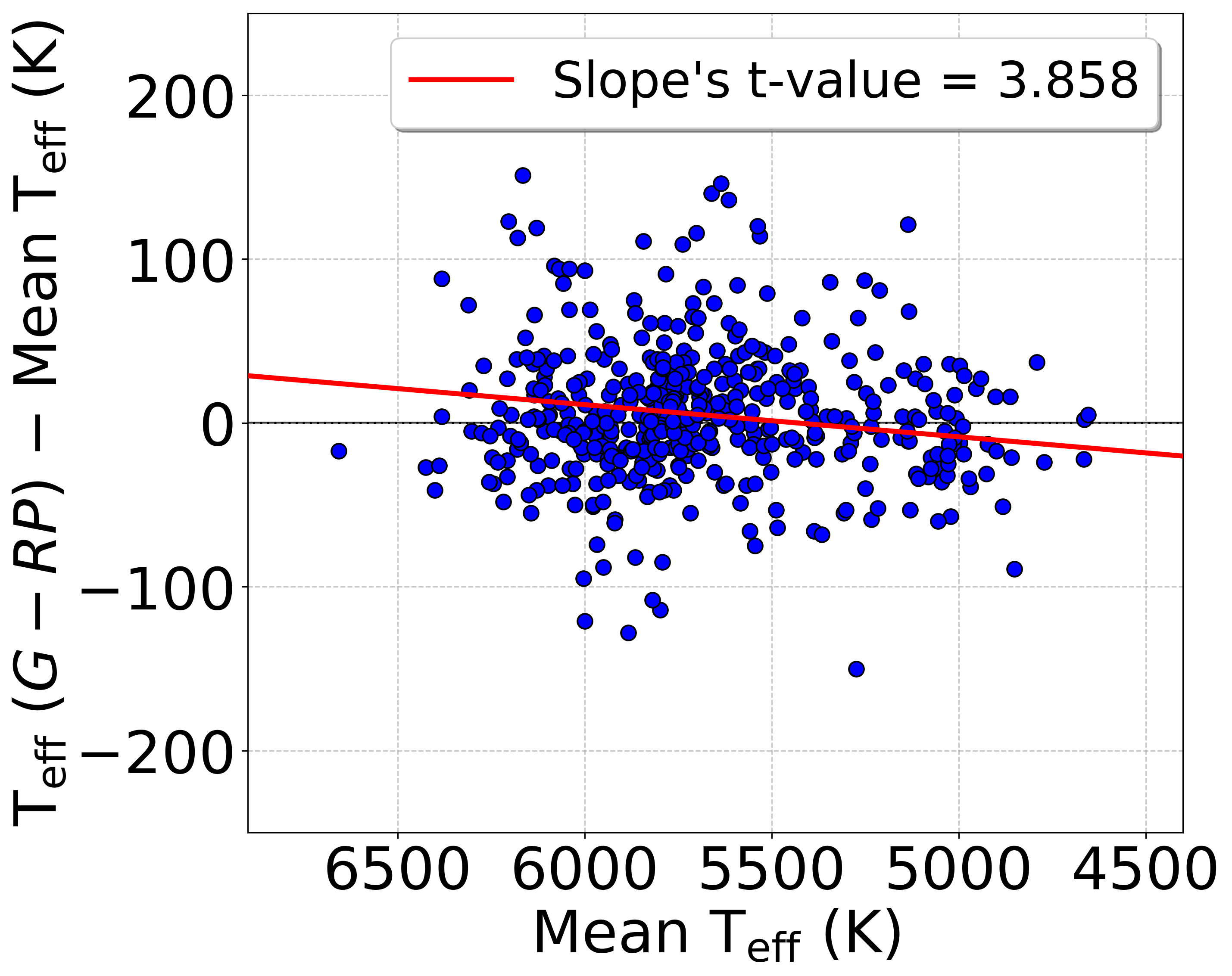} 
    \caption{}
    \vspace{2ex}
  \end{subfigure} 
  \label{fig7}
  \begin{subfigure}[b]{0.49\linewidth}
    \centering
    \includegraphics[width=1\linewidth]{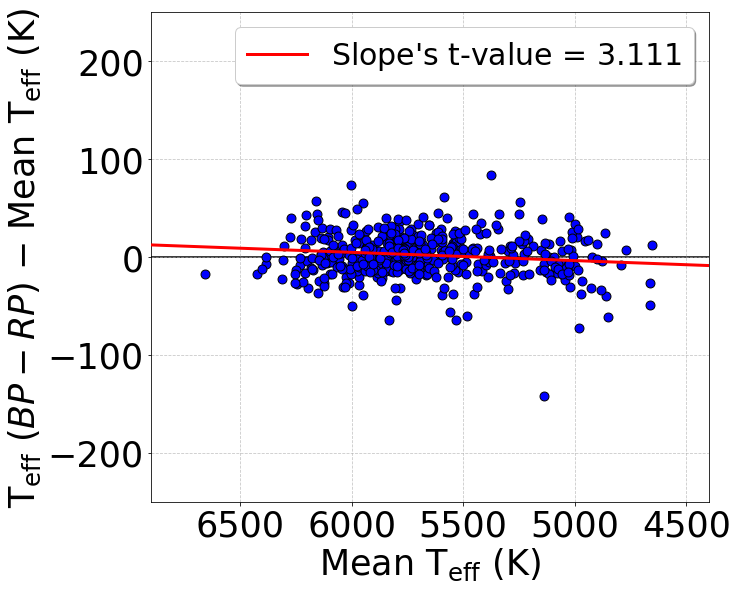} 
    \caption{}
  \end{subfigure}
  \hspace{1ex}
  \begin{subfigure}[b]{0.49\linewidth}
    \centering
    \includegraphics[width=1\linewidth]{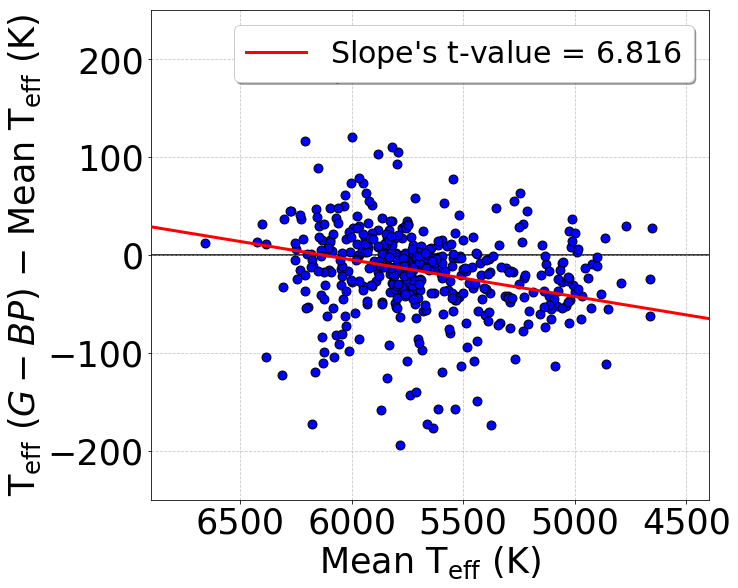} 
    \caption{}
  \end{subfigure} 
  \caption{Behavior of the individual effective temperatures calculated through each colour with respect to the mean value of the parameter. The red line indicates the best fit, and the slope's t-value is also given.}
  \label{fig:teff_trendings} 
\end{figure}

We note the presence of significant trends of the effective temperatures given by the different colours {\it versus} the mean \teff values, with absolute t-values of the slopes of the best-fitting lines higher than 2 (shown in Fig. \ref{fig:teff_trendings}), except $b-y$. We performed some tests removing the individual temperatures with the largest trends (owed to the $B_T - V_T$ and $G-BP$ colours) from the mean. We observed an increase in the t-values of the trend slopes for the remaining colours, while the mean error or the mean T$_{\text{eff}}$ distribution was sustained at 38 K. Thus, we maintained all the individual effective temperatures in the mean calculation as they stand. We suggest that deeper investigations into these trends be considered in future works.

\subsection{Setting \feh to the same scale}

To try and mitigate the inevitable heterogeneity of \teff and \feh values compiled from a large variety of papers, we devised a correction based on the difference between the published \teff value and our adopted photometric \teff (T$_{\text{eff,phot}} = \bar{T}$). Offsets in derived \feh values as a function of varying the adopted \teff in model atmosphere analyses of equivalent widths of Fe I and Fe II are well-known and coherent: we compile some of them, spanning 30 years of literature, in Table \ref{tab:feh_correction}, in the format $\Delta$[Fe/H]/$\Delta$T$_{\text{eff}}$.

\begin{table}
    \centering
    \caption{Relation between offsets in effective temperature and the resulting offsets in \feh, taken from representative literature.}
    \begin{tabular}{c|c}
    \hline
    \hline
    Reference  &  $\Delta$[Fe/H]/$\Delta$T$_{\text{eff}}$ \\
    \hline 
    \cite{clegg1981carbon}  & 0.07 dex/100 K \\
    \cite{steenbock1983barium} & 0.06 dex/100 K \\
    \cite{cayrel1989search} & 0.04 dex/100 K \\
    \cite{zhao1991abundances} & 0.06 dex/100 K \\
    \cite{da2012accurate} & 0.06 dex/100 K \\
    \hline
    \end{tabular}
    \label{tab:feh_correction}
\end{table}

We adopted the representative value of 0.06 dex/100 K to transform the compiled values of [Fe/H], following the equation
\begin{equation}
    \frac{\Delta\text{[Fe/H]}}{\Delta \text{T}_\text{{eff}}} = \frac{0.06 \text{ dex}}{100 \text{ K}} \ ,
    \label{eq: corr_feh}
\end{equation}
where $\Delta\text{[Fe/H]}$ is the correction to be applied and
\begin{equation}
    \Delta \text{T}_\text{{eff}} \equiv \text{T}_{\text{eff,phot}} - \text{T}_{\text{eff,lit}} \ ,
\end{equation}
where T$_{\rm eff,lit}$ stands for the published values.

The sign of the correction refers to the spectroscopic method of deriving \feh. Whenever an analysis adopts, say, too low a \teff as compared to our photometric $\text{T}_{\text{eff}}$, a lower \feh is necessary to explain the observed Fe line intensities, and we introduce a positive correction for \feh (and vice-versa). This procedure should reduce the heterogeneity of our \feh scale: the correction values are tightly distributed between $-$0.10 and $+$0.10 dex, with very few values outside this range, and the mean value is positive, suggesting that, in bulk, the \teff scale adopted by spectroscopic analyses is only slightly cooler than the one from \cite{casagrande2010absolutely} by $\sim$20 K. However, while we could not establish any significant slope in the $\text{T}_{\text{eff,phot}} - \text{T}_{\text{eff,lit}}$ {\it versus} $\text{T}_{\text{eff,phot}}$ regression, there is a significant 70 K scatter between the two scales: it is not uncommon for the \teff difference between the literature and our $\text{T}_{\text{eff,phot}}$ to reach 100 K.

We highlight that the program stars that are members of the young open clusters and stellar associations did not have their metallicities corrected. We chose instead to adopt a collective value of [Fe/H] for each cluster/association from the recent literature (see Table \ref{tab:met_groups}).

\begin{table}
    \centering
    \caption{Metallicities adopted for young associations and clusters. For the only star of the Beta-Pictoris association (HD 35850), we adopted the individual value of [M/H] $= -0.02$ dex of general metallicity \citep{gray2006contributions}.}
    \begin{tabular}{c|c|c}
    \hline
    \hline
     Cluster/Association  &  [Fe/H] & Reference \\
     \hline 
     Hyades & {$+0.18 \pm 0.03$} & \cite{dutra2016consistent}\\
     Pleiades & $+0.01 \pm 0.02$ & \cite{schuler2010fe} \\
     Tucana-Horologium & $-0.06 \pm 0.09$ & \cite{almeida2009search} \\
     \hline
    \end{tabular}
    \label{tab:met_groups}
\end{table}

There were also 41 stars for which we could not find published values of [Fe/H] from spectroscopic analyses. For these cases, we estimated the metallicity from the calibration of \cite{holmberg2007geneva}, based on the colour $b-y$ and the Str\"omgren indexes $c_1$ and $m_1$. A trend was found, along with a zero point offset, while comparing the photometric metallicities (here called [Fe/H]$_\text{phot}$) and the literature ones ([Fe/H]$_\text{lit}$), using all stars with both determinations. We thus added a correction to the values of [Fe/H]$_\text{phot}$ for these 41 stars, following the equation of the best fitting line to the points of the [Fe/H]$_\text{lit}$ $-$ [Fe/H]$_\text{phot}$ \textit{vs} [Fe/H]$_\text{phot}$ plot,
\begin{equation}
    \text{[Fe/H]}_\text{corr} = 0.0178 \times \text{[Fe/H]}_\text{phot} + 0.0868 \ ,
\end{equation}
where $\text{[Fe/H]}_\text{corr}$ is the correction to be summed to [Fe/H]$_\text{phot}$. 


\subsection{Stellar masses, surface gravities and isochronal ages}

The stellar masses and ages were determined using evolutionary tracks and isochrones from the PARSEC suite \citep{bressan2012parsec}. We used a \textit{python} script that automated the process, reading the effective temperature and the bolometric absolute magnitude, calculated using the visual apparent magnitude from HIPPARCOS \citep{van2007validation} and the Gaia DR2 parallaxes, along with the bolometric correction of \cite{flower1996transformations}, as updated by \cite{torres2010use}. The code determines the probability density function for each parameter of interest from a Bayesian formalism \citep{almeida2023chemodynamical} considering the individual uncertainties of the stellar parameters. The code also classifies each star in terms of its probable evolutionary status, as a dwarf, subgiant, or giant.


The solar properties set the scale and zero point of the stellar ages and masses. We determined the zero point correction that brings the PARSEC evolutionary tracks and isochrones to full agreement with the solar mass, luminosity, and age; that is, the present solar T$_{\rm eff}$ and luminosity are recovered for the correct values of mass and age: we adopt $4.57 \times 10^9$ years from \cite{connelly2012absolute}.  The respective corrective values of effective temperature and bolometric absolute magnitude are $+111$ K and $-0.068$ dex, and the grid of evolutionary tracks and isochrones was shifted by these values.

From the bolometric absolute magnitudes, we also calculated the luminosities (L/L$_\odot$) and, in sequence, the stellar radii (R/R$_\odot$) through the Stefan-Boltzmann law. From the stellar mass and radius, we determined the evolutionary surface gravities ($\log{g}$), which we adopt as our best values.

The uncertainties of the absolute bolometric magnitudes, luminosities, radii, and surface gravities were estimated using the usual propagation of errors formula, assuming independence between the variables. Finally, for metallicity \feh, we adopt (somewhat arbitrarily and conservatively) an error of 0.10 dex: for the stars for which only photometric determinations of \feh from the calibration of \cite{holmberg2007geneva}, we adopted 0.15 dex as the error.

\subsection{Second round of iterations for \teff and \feh}
\label{subsec: second round}
Our determination of evolutionary surface gravities for the whole sample allows the further improvement of the accuracy of the \teff and \feh scales. We recomputed the average photometric \teff with the corrected \feh values and the evolutionary $\log{g}$ as input. The whole cycle is summarized as follows:

1) Calculation of the initial T$_{\text{eff,phot}}$, employing the colours plus the metallicities and surface gravities compiled from the literature;

2) Comparison of the T$_{\text{eff,phot}}$ with the one employed by the authors from which we compiled \feh and $\log{g}$, after which we apply the first \feh correction;

3) Recomputation of T$_{\text{eff,phot}}$ employing the corrected \feh values;

4) Determination of stellar bolometric magnitudes, luminosities, masses, and radii employing the corrected \teff and \feh values: determination of evolutionary surface gravities;

5) Recomputation of the T$_{\text{eff,phot}}$ using the values of metallicity and surface gravity determined in items 2 and 4, respectively, closing the first cycle of iterations;

6) Comparison of the iterated T$_{\text{eff,phot}}$ with the original \teff from the literature (T$_{\text{eff,lit}}$), and application of the second (and last) \feh correction;

7) Calculation of the T$_{\text{eff,phot}}$ from the final \feh values and the evolutionary $\log{g}$ determined in item 4;

8) Determination of the final values of stellar bolometric magnitudes, luminosities, masses, and radii from the new T$_{\text{eff,phot}}$ and \feh values: calculation of the final values of evolutionary surface gravity;

9) Calculation of the final T$_{\text{eff,phot}}$ using the final [Fe/H] values (item 6) and the final evolutionary $\log{g}$  values (item 8), closing the second cycle of iterations.

After the second iteration, we observed full convergence for all parameters; the maximum variation of effective temperature between items 7 and 9 is only 1 K. We present all stellar parameters here determined in Tables \ref{tab:atmospheric_parameters} and \ref{tab:other_parameters} with their errors, for each star. In short, the mean error of effective temperature remained 38 K; the $\log{g}$, mass, and radius mean errors are, respectively, 0.05 dex, 0.05 M$_\odot$ and 0.03 R$_\odot$. Isochronal ages were included in Table \ref{tab:other_parameters} only if their determination carried a mean error lower than 1.0 Gyr, which we consider the internal precision limit for building the age-activity relation (Sec. \ref{sec5}).

Henceforth, T$_\text{eff}$, [Fe/H], and $\log{g}$ refer exclusively to the final set of iterated atmospheric parameters, T$_{\text{eff,phot}}$ and \feh corrected for the T$_{\text{eff,phot}}$ scale: all evolutionary parameters are consistent with this adopted \teff and \feh scale. In a final validation test, we compared our atmospheric parameters with those from the Gaia FGK Benchmark Stars \citep{jofre2018gaia} and the TITANS metal-poor reference stars \citep{giribaldi2021titans}, for all stars in common (13 stars and four stars, respectively). For the Gaia benchmark stars, the mean difference (this work $-$ theirs) was $13 \pm 45$ K in T$_\text{eff}$, $0.01 \pm 0.07$ dex in $\log{g}$ and $0.01 \pm 0.05$ dex in [Fe/H], without any trend of the differences with the parameters themselves. For the TITANS stars, the mean difference was $8 \pm 32$ K in T$_\text{eff}$, $0.02 \pm 0.02$ dex in $\log{g}$, and $0.06 \pm 0.07$ dex in \feh. Our atmospheric parameters are thus in line with recent determinations.

\begin{table*}
  \centering
   \caption{Atmospheric parameters collected from the literature and the final values used in this paper. The first column indicates the star identification (HD number). The second to fourth columns show the atmospheric parameters compiled from the literature, with the reference specified in the fifth column. Unavailable parameters are indicated by "--". Stars for which no spectroscopic metallicity was available had their \feh estimated from the \citet{holmberg2007geneva} calibration, as indicated in the "Reference" column with "H". T$_{\text{eff,phot}}$ and its uncertainty, corrected \feh and evolutionary $\log{g}$ computed in the present work are, respectively, shown in the last four columns. The full table, including the full literature list for the atmospheric parameters, is available only online.} 
  \begin{tabular}{p{1.0cm}p{1.3cm}p{1.3cm}p{1.3cm}p{1.8cm}p{1.3cm}p{1.3cm}p{1.3cm}p{1.3cm}p{1.3cm}}
 \hline
 \hline
\multicolumn{1}{c}{} & \multicolumn{3}{c}{Literature} & \multicolumn{1}{c}{} & \multicolumn{5}{c}{Adopted in this work} \\ 
\hline
HD & T$_{\text{eff}}$ (K) & [Fe/H] & $\log{g}$ & Reference & T$_{\text{eff,phot}}$ (K) & $\sigma$ (K) & [Fe/H] & $\log{g}$ & $\sigma$\\
\hline
105 & 6126 & $ +0.08 $ & 4.65 & 39 & 5960 & 27 & $ -0.06 $ & 4.42 & 0.04 \\
166 & 5465 & $ +0.14 $ & 4.53 & 29 & 5521 & 21 & $ +0.17 $ & 4.55 & 0.03 \\
1237 & 5541 & $ +0.12 $ & 4.54 & 29 & 5480 & 28 & $ +0.08 $ & 4.51 & 0.05 \\
1461 & 5724 & $ +0.17 $ & 4.35 & 29 & 5743 & 13 & $ +0.18 $ & 4.37 & 0.04 \\
1466 & 6135 & $ -0.06 $ & 4.39 & 19 & 6139 & 41 & $ -0.06 $ & 4.40 & 0.05 \\
1581 & 5922 & $ -0.21 $ & 4.35 & 29 & 6009 & 70 & $ -0.16 $ & 4.43 & 0.05 \\
1835 & 5817 & $ +0.21 $ & 4.46 & 29 & 5772 & 23 & $ +0.18 $ & 4.44 & 0.04 \\
2151 & 5816 & $ -0.12 $ & 3.95 & 29 & 5893 & 65 & $ -0.07 $ & 4.00 & 0.04 \\
3047 & -- & $ +0.02 $ & -- & H & 5901 & 22 & $ +0.02 $ & 3.98 & 0.08 \\
3443 & 5501 & $ -0.21 $ & 4.28 & 18 & 5551 & 61 & $ -0.18 $ & 4.23 & 0.05 \\
... & ... & ... & ... & ... & ... & ... & ... & ... & ... \\
\hline
  \end{tabular}
  \label{tab:atmospheric_parameters}
\end{table*}

\begin{table*}
  \centering
  \caption{Bolometric correction (BC) and magnitude (M$_{\text{bol}}$), luminosity (L/L$_\odot$), radius (R/R$_\odot$), mass (M/M$_\odot$) and age with its respective uncertainties. The inferior and superior age errors are indicated separately; for the mass, the uncertainty was taken as the average between the two values. The lack of ages indicated by "--" represents the cases in which we could not determine the age with the necessary precision (error  $<$ 1 billion yrs.). The last column indicates the stars with ages determined by asteroseismology (*, see Table. \ref{tab:ages}) and stars that are members of young open clusters or stellar associations, whose ages were taken from specific and more detailed works (A, see Table. \ref{tab:group_ages}). If this column is empty, the age was determined ordinarily through isochrones. The full table is available online.}
  \begin{tabular}{p{0.4cm}p{0.85cm}p{0.6cm}p{0.7cm}p{0.7cm}p{0.7cm}p{0.7cm}p{0.7cm}p{0.7cm}p{0.8cm}p{1.2cm}p{1.5cm}p{1.5cm}p{0.5cm}}
  \hline
  \hline
  HD & BC & M$_{\text{bol}}$ & $\sigma_{\text{M}_\text{bol}}$ & L/L$_\odot$ & $\sigma_{\text{L}/\text{L}_\odot}$ &  R/R$_\odot$ & $\sigma_{\text{R}/\text{R}_\odot}$ & M/M$_\odot$ & $\sigma_{\text{M}/\text{M}_\odot}$ & Age (Gyr.) & $\sigma_{\text{Age}}^{\text{INF}}$ (Gyr.)& $\sigma_{\text{Age}}^{\text{SUP}}$ (Gyr.) & Note \\ 
  \hline
105 & $ -0.051 $ & 4.513 & 0.008 & 1.222 & 0.009 & 1.037 & 0.010 & 1.036 & 0.050 & 0.045 & 0.004 & 0.004 & A \\
166 & $ -0.133 $ & 5.241 & 0.007 & 0.625 & 0.004 & 0.864 & 0.007 & 0.963 & 0.021 & -- & -- & -- \\
1237 & $ -0.144 $ & 5.223 & 0.009 & 0.635 & 0.005 & 0.884 & 0.010 & 0.933 & 0.053 & -- & -- & -- \\
1461 & $ -0.086 $ & 4.531 & 0.006 & 1.201 & 0.007 & 1.107 & 0.006 & 1.049 & 0.065 & -- & -- & -- \\
1466 & $ -0.027 $ & 4.267 & 0.007 & 1.532 & 0.011 & 1.094 & 0.015 & 1.099 & 0.043 & 0.045 & 0.004 & 0.004 & A \\
1581 & $ -0.044 $ & 4.530 & 0.012 & 1.202 & 0.014 & 1.011 & 0.024 & 1.008 & 0.043 & -- & -- & -- \\
1835 & $ -0.081 $ & 4.666 & 0.007 & 1.061 & 0.007 & 1.030 & 0.009 & 1.067 & 0.041 & 0.625 & 0.050 & 0.050 & A \\
2151 & $ -0.061 $ & 3.396 & 0.011 & 3.417 & 0.036 & 1.773 & 0.040 & 1.139 & 0.049 & 6.400 & 0.560 & 0.560 & * \\
3047 & $ -0.059 $ & 3.246 & 0.053 & 3.921 & 0.190 & 1.894 & 0.048 & 1.237 & 0.053 & 4.880 & 0.611 & 0.476 &  \\
3443 & $ -0.126 $ & 4.506 & 0.064 & 1.229 & 0.072 & 1.199 & 0.044 & 0.893 & 0.023 & 12.70 & 1.130 & 0.640 &  \\
... & ... & ... & ... & ... & ... & ... & ... & ... & ... & ... & ... & ... & ... \\
\hline
  \end{tabular}
  \label{tab:other_parameters}
\end{table*}

  \label{fig:comp_logg_iteraction}

\section{- Determination of the chromospheric absolute fluxes}
\label{sec4}

In this section, we derive the total absolute fluxes at the H$\alpha$ line core, in units of erg $\text{cm}^{-2} \text{ s}^{-1}$, composed of the photospheric part, which is largely dominant, plus the chromospheric component. We then describe the procedure to correct for photospheric flux and obtain the purely chromospheric component.

\subsection{- Theoretical absolute flux calibration}
We used theoretical stellar atmosphere models from the latest version of the NMARCS suite \citep{gustafsson2008grid} to determine the total absolute fluxes. The LTE models cover the following range of the atmospheric parameters: $4800 <$ \teff (K) $< 6400$, $-1.0 <$ \feh $< +0.4$ and $3.4 <$ $\log{g}$ $< 4.6$, spaced, respectively, by 200 K, 0.2 dex and 0.2 dex.

The core of the H$\alpha$ line is formed very high in the stellar atmospheres \citep{vernazza1981structure,leenaarts2012formation}, and the LTE approximation is not valid. We employed a semi-empirical approach, relating the observed flux in selected continuum windows in the spectra (for which the theoretical flux is accurate) to the observed fluxes at the H$\alpha$ line core. We thus quantify the latter following the equation
\begin{equation}
    \frac{F_{H\alpha}}{F_{ref}} = \frac{f_{H\alpha}}{f_{ref}} \times \Delta \lambda \ ,
    \label{eq:fluxoAparAbs}
\end{equation}
where $F_{H\alpha}$ is the real absolute flux in the line (the variable of interest), $F_{ref}$ is the real absolute flux in the continuum region, given by the theoretical models, $f_{H\alpha}$ and $f_{ref}$ are the apparent fluxes, measured directly off the spectra, and $\Delta \lambda$ the bandwidth used for the continuum regions. This procedure is superior to that employed by \cite{pasquini1991h} and LPM05 since the total flux at the line core is referenced to accurate line-free continuum fluxes, not far in wavelength to the H$\alpha$ line core and calculated with internally precise atmospheric parameters determined for each star, explicitly taking into account the \feh and $\log{g}$ dependence.

\begin{figure}
    \centering
    \includegraphics[width=\columnwidth]{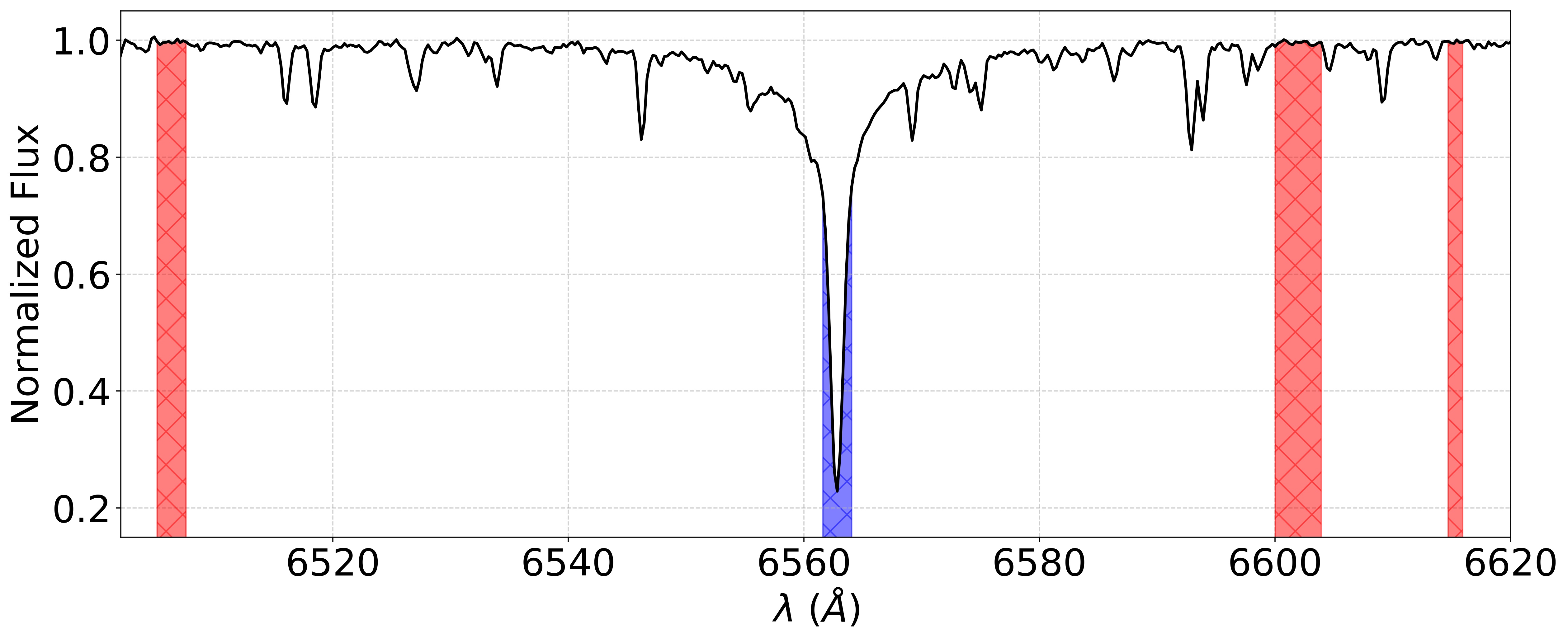}
    \caption{The three continuum regions (in red) and the H${\alpha}$ region (in blue) in which we measured, respectively, $f_{H\alpha}$ and $f_{ref}$, in the spectrum of the solar twin star HD 146233 at low resolution, R = 11\,000.}
    \label{fig:hd146233_m125_n1_en}
\end{figure}

The continuum regions used were three: (1) 6504.95 - 6507.55 \AA, (2) 6599.96 -  6604.25 \AA, \ and (3) 6614.49 - 6616.15 \AA. The apparent fluxes of these regions are highlighted in Fig. \ref{fig:hd146233_m125_n1_en}, as well as the 
apparent flux to be measured in the H${\alpha}$ line. Following equation \ref{eq:fluxoAparAbs}, we obtained three estimates for $F_{H\alpha}$, and a simple mean of these values established the final value.

We computed absolute fluxes in the continuum regions ($F_{ref}$) tailored to the individual atmospheric parameters of the program stars. The fluxes were estimated by a cubic multiparametric regression on T$_\text{eff}$, [Fe/H] and $\log{g}$ of the grid models, obtaining $F_{ref} = F_{ref}(\text{T}_\text{eff},\text{[Fe/H]},\text{$\log{g}$})$ representations for each continuum region. The mean residual of the $F_{ref} = F_{ref}(\text{T}_\text{eff},\text{[Fe/H]},\text{$\log{g}$})$ regression is $\sigma_{reg} = 8 \times 10^3$ erg cm$^{-2}$ s$^{-1}$, a negligible value compared with typical values of the chromospheric fluxes, as will be shown later. 


\subsection{Determination of the absolute flux in the H$\alpha$ line core}
\begin{figure}
  \begin{subfigure}[b]{0.493\linewidth}
    \centering
    \includegraphics[width=1\linewidth]{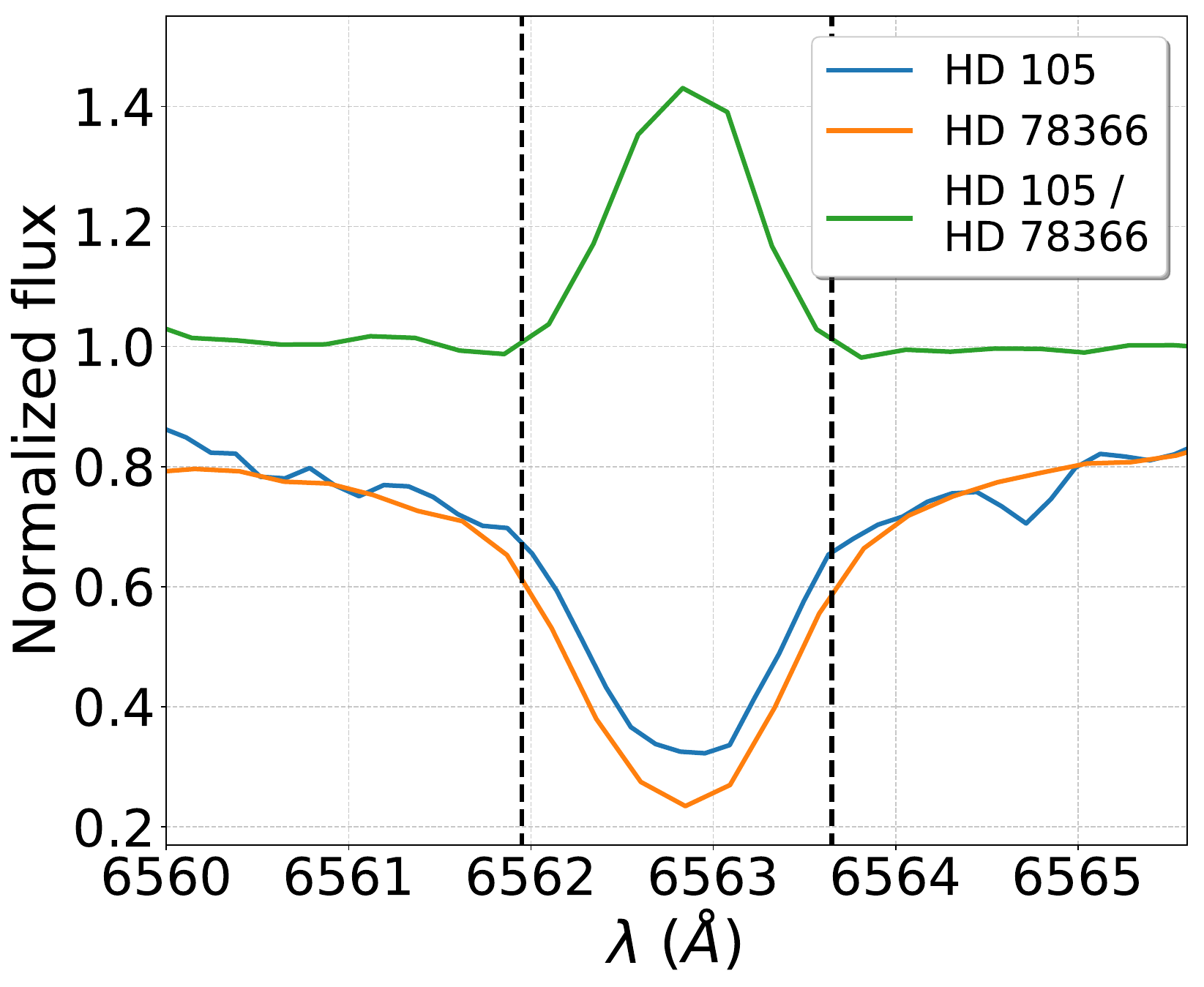} 
    \caption{}
  \end{subfigure}
  \begin{subfigure}[b]{0.507\linewidth}
    \centering
    \includegraphics[width=1\linewidth]{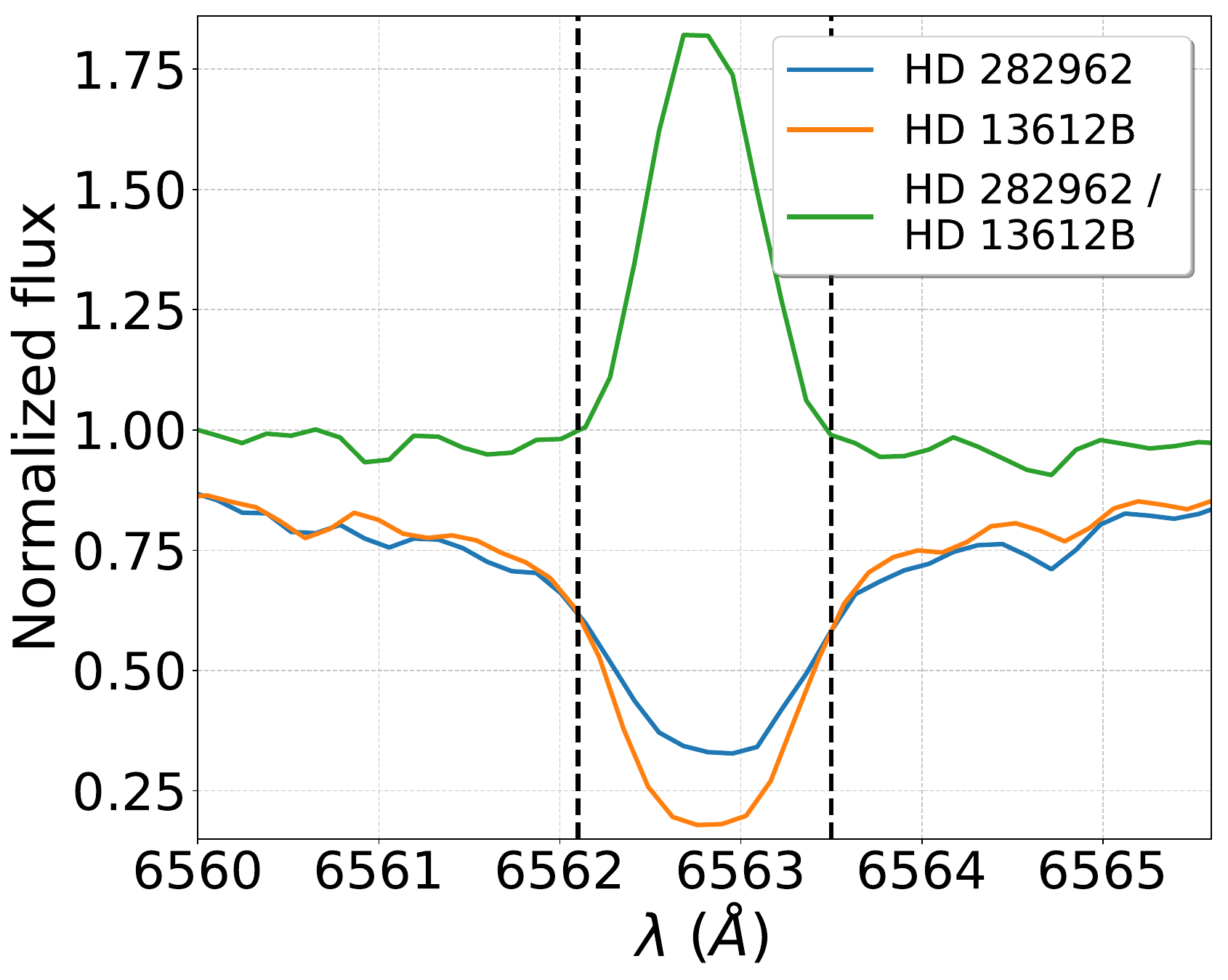} 
    \caption{}
  \end{subfigure} 
  \caption{Normalized flux ratio spectra (in green) for pairs of active-inactive stars (in each case with identical atmospheric parameters but widely different levels of chromospheric emission) showcased along with the individual spectra (in orange and blue), for data of (a) low resolution, R = 11\,000 and (b) high resolution, R = 30\,000. The non-ratioed spectra have been arbitrarily displaced downward. Black dashed vertical lines bracket the characteristic bandwidth we adopted to include all of the chromospheric contribution.}
  \label{fig:ratio_spectra}
\end{figure}

The next step is determining the ideal bandwidth to measure the flux around the center of the H$\alpha$ line so we can quantify $f_{H\alpha}$ and, finally, $F_{H\alpha}$. In the H$\alpha$ line, the chromospheric emission is given by a smooth filling on its core, as seen in Fig. \ref{fig:ratio_spectra}, lacking the well-known structure of emission and self-absorption of the Ca II H+K lines. The stars highlighted in Fig. \ref{fig:ratio_spectra} have, within each comparison, very similar atmospheric parameters, and therefore, the observed flux difference in the spectra is assumed to be entirely due to a difference in chromospheric fill-in emission. Essentially, we presume the photospheric component of the flux to be separable from the chromospheric one and accountable by the theoretical representation of the model atmospheres solely as a function of T$_\text{eff}$, [Fe/H] and $\log{g}$.

The bandwidth to integrate the line core flux needs to be wide enough to include all of the chromospheric emission, a problem already reviewed at length in the literature \citep[LPM05]{pasquini1991h}. Ensuring that the chromospheric fill-in is fully accounted for is a more fundamental problem than allowing excess, unwanted purely photospheric flux to be included in the integration, since the subtraction procedure will merely eliminate this excess. Our concern is thus allowing the bandwidth to be as wide as necessary. For this analysis, we selected many pairs of stars with different levels of magnetic activity but with nearly equivalent atmospheric parameters and considered within the same resolution group of observed spectra. For each pair, we considered the spectrum of the more active star ratioed to the spectrum of the more inactive star. Two examples are seen in Fig. \ref{fig:ratio_spectra}. The ideal bandwidth is limited to the wavelength range where the flux ratio differs from unity, including all of the prominence peak centered at the line core.

It is apparent in Fig. \ref{fig:ratio_spectra} that the ``ideal'' bandwidth for the different spectral resolutions, $R \sim 11\,000$ and $R \sim 30\,000$, is not the same, the lower resolution spectra requiring, as expected, a larger bandwidth than the spectra with higher resolution. \cite{herbig1985chromospheric}, \cite{pasquini1991h} and LPM05 have all converged to the same value, using high-resolution spectra: $\Delta \lambda_{H\alpha} = 1.7$ \AA . We defined $\Delta \lambda_{H\alpha} = 2.4$ and 1.8 \AA \, respectively, for R = 11\,000 and R = 30\,000 spectra, essentially because these were the larger values set by data with good S/N, after we analyzed all possible ratio spectra of stars sharing the same T$_\text{eff}$, [Fe/H] and $\log{g}$, particularly scrutinizing ratio spectra of dwarf/subgiant pairs in an attempt to ascertain that the chromospheric component in the wider cores of evolved stars is included in the flux computation. Evolved stars show unequivocally wider H$\alpha$ doppler-cores than less evolved stars at the same T$_{\text{eff}}$. Possible trends of the bandwidths with the atmospheric parameters of the stars were investigated but deemed inconclusive.

Absolute fluxes at the H$\alpha$ line core, averaged for the three continuum windows, were computed from Eq.\ref{eq:fluxoAparAbs}. The next step is to homogenize data from the different spectral resolutions, computed from different $\Delta \lambda_{H\alpha}$ bandwidths, to the same scale. The larger $R \sim 30\,000$ subsample was chosen as the standard. We plotted the absolute fluxes for the 62 stars with spectra at both resolutions and derived a simple linear regression. The result is a very tight correlation (R = 0.955)
\begin{equation}
    F_{30000} = 0.619 \times F_{11000} - 9.04 \times 10^4 \ ,
    \label{eq:conversion_fluxes}
\end{equation}
where $F_{30000}$ and $F_{11000}$ are, respectively, the absolute fluxes for $R \sim 30\,000$ and $R \sim 11\,000$. The uncertainty associated with the conversion is a significant one for the total error budget: it is set at the standard deviation of the regression, $\sigma = 1.47 \times 10^5$ $\text{ergs cm}^{-2} \text{ s}^{-1}$.

All fluxes are thus homogenized; we no longer distinguish between the two distinct resolving powers and average the flux values for the stars that have more than one observation. At least two observations are available for 124 stars (24 \% of the sample).

\subsection{Photospheric correction and final chromospheric absolute fluxes}
\begin{figure}
    \centering
    \includegraphics[width=7.5cm]{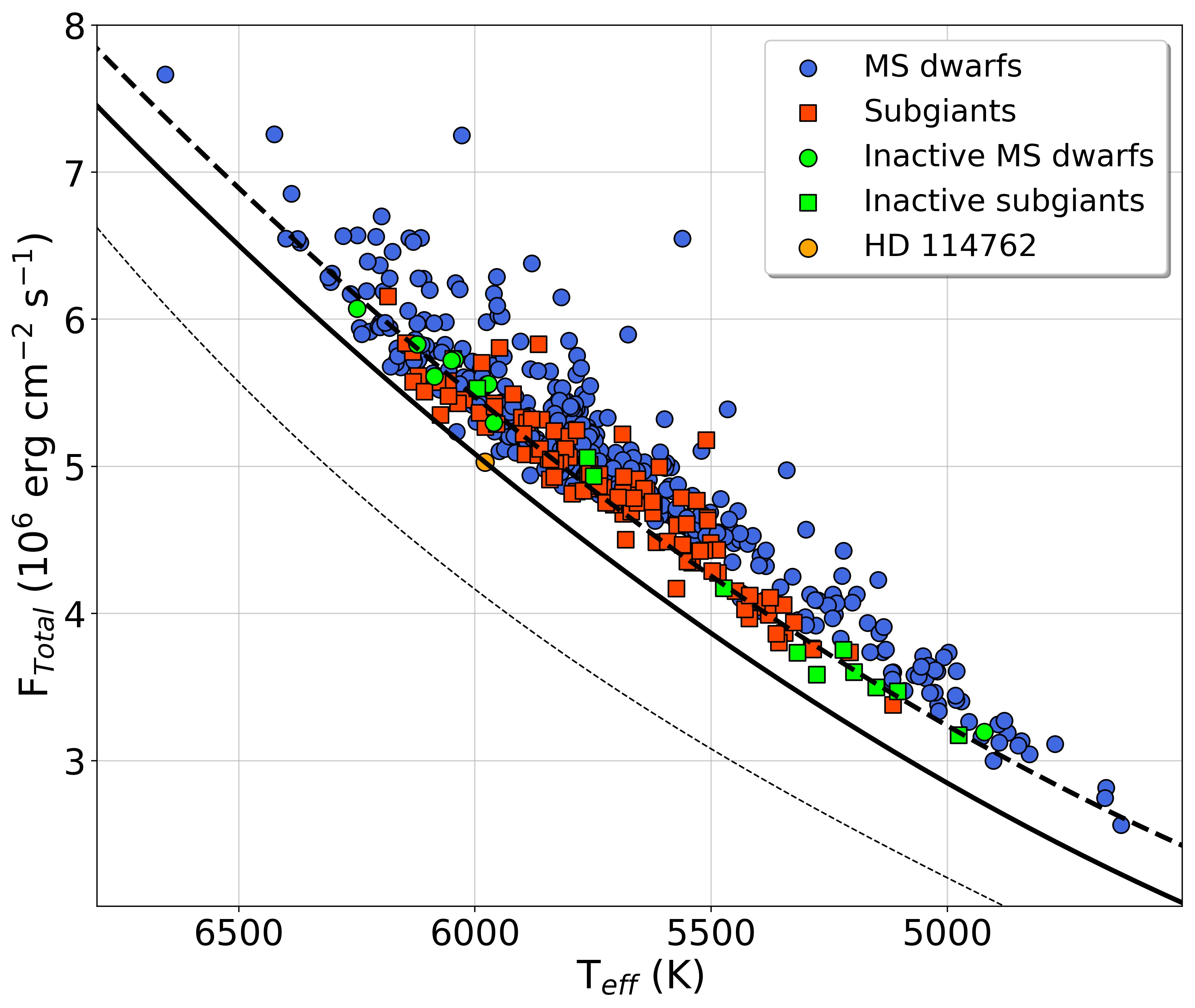}
    \caption{Absolute total fluxes as a function of \teff for the 511 sample stars, dwarfs (blue circles) and subgiants (red squares). The black dashed line indicates the best fitting line to the H$\alpha$ total fluxes for the most inactive stars considering the values of log($R'_{HK}$) (in green); the black continuous line represents this same fitting line vertically shifted to fit the star HD 114762 (orange circle). For illustration, we also show as the thin dashed line the photospheric correction (scale shifted) applied to the $\lambda$8498 \ion{Ca}{ii} infrared triplet by \citet{lorenzo2016fine}.}
    \label{fig:envelope}
\end{figure}

We obtained the final chromospheric absolute fluxes by removing the purely photospheric component, which we deem as dependent exclusively on the atmospheric parameters, T$_\text{eff}$, [Fe/H], and $\log{g}$. In Fig. \ref{fig:envelope}, we plot the total absolute fluxes for the whole sample as a function of T$_{\text{eff}}$. We adopt the same procedure as LPM05 and define the photospheric correction as the total absolute flux of the least active star at a given \teff, under the hypothesis that the lowest flux corresponds to a star with null, or, more rigorously, minimum, chromospheric activity. Total fluxes are highly correlated to T$_\text{eff}$: we employed this fact to our advantage by estimating the photospheric contribution. A second-order polynomial was fitted to the 20 most inactive stars using the log($R'_{HK}$) indicator (Sect. \ref{sec5}) for what we assume to represent the general trend of the photospheric contribution. This function is then displaced vertically to fit the least possible active star in the H$\alpha$ fluxes, in this case, one single star, HD 114762. This envelope of minimal activity, defined for each T$_\text{eff}$, is subtracted from the total stellar fluxes, providing the final purely chromospheric fluxes. The equation that describes the photospheric correction is
\begin{equation}
    \text{F}_{\text{phot}} (\text{erg } \text{cm}^{-2} \text{ s}^{-1}) = 0.4 \cdot \text{T}_\text{eff}^2 - 2167 \cdot \text{T}_\text{eff} + 3.676 \times 10^6\ .
\end{equation}

Similarly to LPM05 and \cite{pasquini1991h}, we find (Fig.\ref{fig:envelope}) that the lowest H$\alpha$ fluxes are systematically populated by subgiant stars, even though the photospheric correction envelope is defined by the dwarf star HD 114762, which is arbitrarily set to possess null chromospheric flux. HD 114762 is a low-mass, low-metallicity (M/M$_\odot$ $\sim$ 0.83 and \feh $\sim$ -0.7), very evolved star (R/R$_\odot$ $\sim$ 1.17, L/L$_\odot$ $\sim$ 1.59, $\log{g}$ $\sim$ 4.2) star for which we could not determine the isochronal age. We note that this procedure sets the zero point of the chromospheric flux scale but does not influence the age scale, which is independently set in Sect. \ref{sec3}. All stars with total absolute flux values close to the envelope of photospheric correction will automatically have very uncertain chromospheric fluxes, and no meaningful chromospheric ages will be attributable to them.

This empirical, heavily sample-dependent approach to defining the photospheric correction is objectionable on several accounts. It differs fundamentally from that of \cite{pasquini1991h}, who compared H$\alpha$ chromospheric fluxes and F'$_\text{K}$ fluxes (the latter corresponding to absolute chromospheric fluxes for the \ion{Ca}{ii} K line) and forced the former to go to zero along with the latter. This somewhat more physical treatment can also be criticized because H$\alpha$ and the \ion{Ca}{ii} H+K lines are formed at slightly different depths within stellar chromospheres. These lines reflect different formation physics and are characteristic of different atmospheric levels \citep{schoolman1972formation,vernazza1981structure}.

In the present work, we accept the arbitrary nature of the photospheric correction defined in Fig.\ref{fig:envelope}, yet we hypothesize that the extension of the sample towards very inactive, subgiant stars, as well as very metal-poor, old stars, in all probability will push the photospheric correction to lower levels and change the procedure we adopt here. It seems reasonable to suppose that for still more evolved, slower rotating stars, even lower chromospheric fluxes will be observed and that no cool star will effectively ever reach null values of chromospheric contribution. There is no consensus in the literature, as yet, on either the presence of basal chromospheric fluxes \citep{schrijver1995basal} or its magnetic and/or acoustic character \citep[][and references therein]{schrijver2023stellar}. Ideally, purely photospheric fluxes at the H$\alpha$ line core should be computed for a full NLTE grid of models. This approach is, however, computationally expensive and falls beyond the scope of this work.

\subsection{Error budget evaluation}

A comprehensive list of uncertainties affecting the final error budget of the chromospheric fluxes is:
\begin{enumerate}
    \item residuals of the cubic regressions to determine the theoretical absolute fluxes in the continuum regions;
    \item uncertainties in the theoretical absolute fluxes of the continuum regions reflected by the errors in \teff, $[$Fe/H$]$ and $\log{g}$;
    \item standard deviation of the H$\alpha$ mean total fluxes averaged for the three continuum regions;
    \item  standard deviation of the mean H$\alpha$ total fluxes, when more than one measurement is available for a star;
    \item standard deviation of the conversion equation between the two flux scales (high and low-resolution spectra), only for fluxes coming from the low resolution spectra, which are the ones converted.
\end{enumerate}

\begin{figure}
    \centering
    \includegraphics[width=\columnwidth]{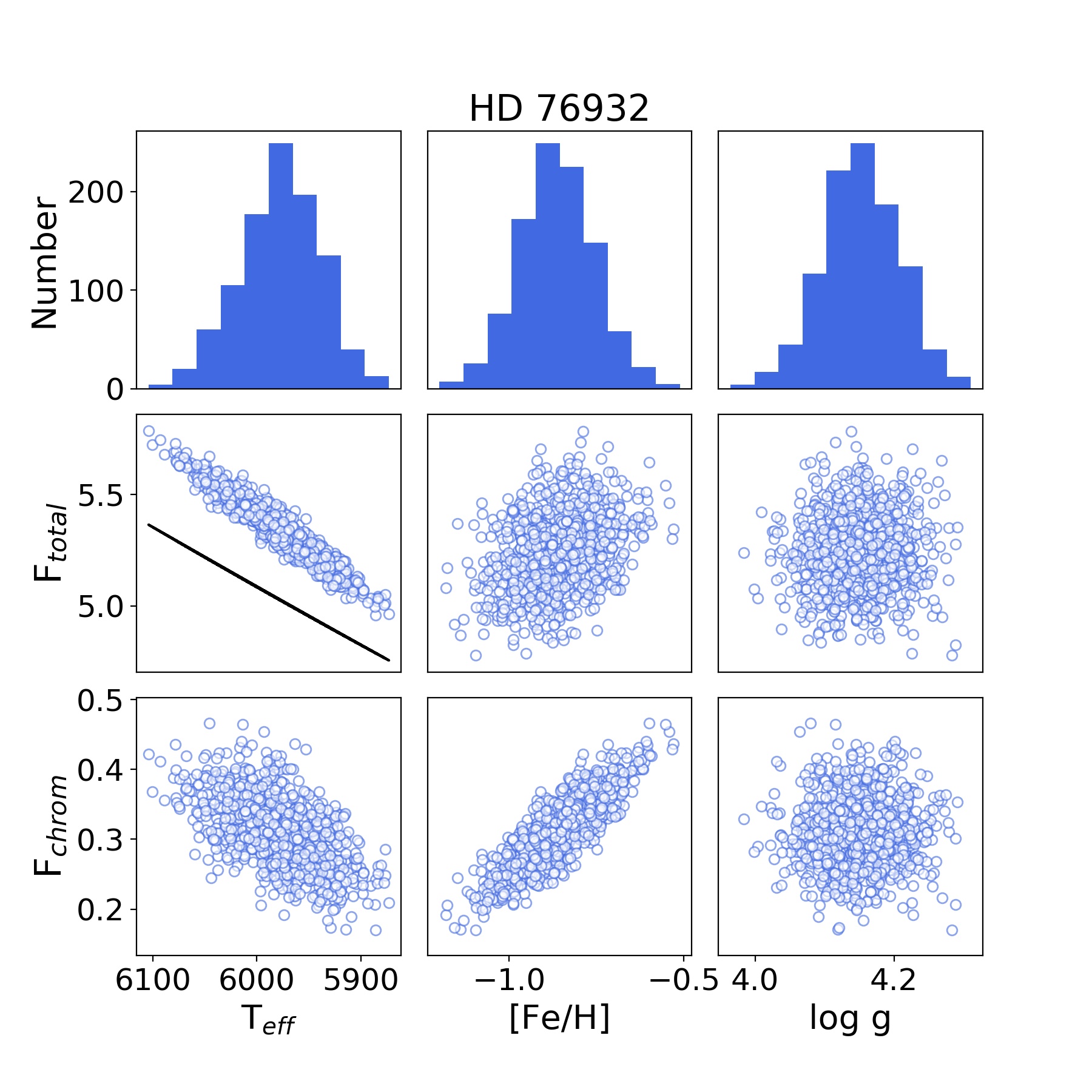}
    \caption{Propagation of the errors of $T_{\rm eff}$, $\log{g}$ and $[$Fe/H$]$ in the total and chromospheric absolute fluxes utilizing Monte Carlo simulations, for the star HD 76932. Upper panels show the result of 10$^{\rm 3}$ simulations around the mean values of the parameters; middle panels show their reflection in the total flux (the line in the left-center panel is the run of photospheric correction); lower panels show their reflection in the chromospheric fluxes.}
    \label{fig:montecarlo}
\end{figure}

We performed Monte Carlo simulations (N = $10^3$) for each star, generating Gaussian distributions of \teff, \feh and $\log{g}$ centered on each star's adopted values and based on their typical uncertainties (respectively, 38 K, 0.10 dex and 0.05 dex) to estimate the errors reflected by these parameters on the calculation of the total fluxes. The corresponding distribution of chromospheric fluxes was calculated by subtracting the photospheric correction defined in subsection 4.3. To illustrate the procedure, we show in Fig. \ref{fig:montecarlo} the distributions generated for HD 76932, a star with the same \teff of HD 114762 (which cannot be used for this evaluation, being defined as having exactly zero chromospheric flux), but slightly more active. The standard deviations of the simulated distributions of chromospheric fluxes (due to each atmospheric parameter) were quadratically added to the uncertainties provided by items i, iii, iv, and v above, using $\sigma^2 = \sum \sigma_i^2$.

Similarly to \cite{lorenzo2016fine} (Paper I) for the chromospheric fluxes in the infrared Ca II triplet, we find a strong correlation of the simulated total fluxes with \teff, which partially translates in a correlation of the chromospheric fluxes with \teff. This can be appreciated in the left-middle and left-lower panels of Fig. \ref{fig:montecarlo}. However, unlike Fig. 12 of \cite{lorenzo2016fine}, the inclination of the distribution of H$\alpha$ simulated total fluxes with \teff differs from the corresponding inclination of the envelope of photospheric correction. The near-perfect cancellation of the photospheric correction errors in the chromospheric triplet line fluxes is not realized for H$\alpha$, and a residual \teff dependence must be considered. Table \ref{tab:fluxes} shows the values of total fluxes and chromospheric fluxes and their errors.

\begin{table}
  \centering
  \caption{Final total and purely chromospheric fluxes for the program stars. The first column is the HD number; the second column is total absolute flux, F$_{\rm total}$; the third column is the chromospheric component, F$_{\rm chrom}$, with its associated uncertainty $\sigma_{\rm Fchrom}$ in the last column. The full table is available online.}
  \begin{tabular}{p{0.7cm}p{1.7cm}p{1.7cm}p{1.7cm}}
 \hline
 \hline
HD & F$_{\text{total}}$ ($\times 10^6$ & F$_{\text{chrom}}$ ($\times 10^6$ & $\sigma_\text{Fchrom}$ ($\times 10^6$\\ 
 & erg $\text{cm}^{-2} \text{ s}^{-1}$) & erg $\text{cm}^{-2} \text{ s}^{-1}$) & erg $\text{cm}^{-2} \text{ s}^{-1}$) \\ 
\hline
105 & 6.1760 & 1.1967 & 0.2028 \\
166 & 5.1051 & 1.1920 & 0.1570 \\
1237 & 4.7784 & 0.9570 & 0.1476 \\
1461 & 5.2141 & 0.7812 & 0.1633 \\
1466 & 6.5522 & 1.0939 & 0.2150 \\
1581 & 5.5336 & 0.4258 & 0.1604 \\
1835 & 5.4911 & 0.9873 & 0.1506 \\
2151 & 5.0826 & 0.2760 & 0.1439 \\
3047 & 5.3314 & 0.5044 & 0.1520 \\
3443 & 4.5854 & 0.6044 & 0.1453 \\
... & ... & ... &  ... \\
\hline
  \end{tabular}
  \label{tab:fluxes}
\end{table}

As a further check on the uncertainties of our measurements, we searched for possible modulations in the H$\alpha$ absolute fluxes for all stars for which we observed more than three spectra on different dates (Table \ref{tab:repeatability}). The mean value $\sigma_{\bar{\rm F}_{\text{total}}}$/$\bar{\rm F}_{\text{total}}$ = 0.0141 $\pm$ 0.0069 testifies to the high internal consistency of our observations (spanning nearly three decades and different detectors) and also that H$\alpha$ is little affected by either stellar cycles or transient phenomena.

\begin{table}
  \centering
  \caption{Total mean absolute fluxes ($\bar{\rm F}_{\text{total}}$) for all objects with n $\geq$ 3 spectra. The first column identifies the star; second column gives the number of available spectra; third and fourth columns, respectively, list the mean absolute flux $\bar{\rm F}_{\text{total}}$ and its absolute uncertainty, $\sigma_{\bar{\rm F}_{\text{total}}}$, in erg $\text{cm}^{-2} \text{ s}^{-1}$. Last column gives the relative error $\sigma_{\bar{\rm F}_{\text{total}}}$/$\bar{\rm F}_{\text{total}}$. The top panel lists the R = 11\,000 low-resolution data, giving raw data, that is, {\bf not} converted to the R = 30\,000 flux scale through Eq. \ref{eq:conversion_fluxes}). The bottom panel lists the R = 30\,000 high-resolution data. Please note that the two flux scales are not interchangeable.}
  
  \begin{tabular}{p{0.9cm}p{0.4cm}p{1.7cm}p{1.7cm}p{1.6cm}}
 \hline
 \hline
HD & N & $\bar{\rm F}_{\text{total}}$ ($\times 10^6$ & $\sigma_{\bar{\rm F}_{\text{total}}}$ ($\times 10^6$ & $\sigma_{\bar{\rm F}_{\text{total}}}$/$\bar{\rm F}_{\text{total}}$ \\ 
 & & erg $\text{cm}^{-2} \text{ s}^{-1}$) & erg $\text{cm}^{-2} \text{ s}^{-1}$) & \\
 \hline
10700 & 3 & 6.7152 & 0.1437 & 0.021 \\
53705 & 3 & 8.4351 & 0.0694 & 0.008 \\
84117 & 3 & 9.7806 & 0.1306 & 0.013 \\
101501 & 3 & 7.5270 & 0.0979 & 0.013 \\
104304 & 5 & 7.7024 & 0.0910 & 0.012 \\
114613 & 4 & 7.9733 & 0.0644 & 0.008 \\
114946 & 4 & 5.4225 & 0.0309 & 0.006 \\
140283 & 3 & 7.6225 & 0.0419 & 0.006 \\
146233 & 9 & 8.6161 & 0.0630 & 0.007 \\
\hline
Sun & 14 & 5.1202 & 0.8169 & 0.016 \\
1581 & 4 & 5.5336 & 1.0930 & 0.020 \\
1835 & 6 & 5.4911 & 0.09251 & 0.017 \\
2151 & 4 & 4.9943 & 0.0536 & 0.011 \\
10700 & 4 & 4.1417 & 0.0177 & 0.004 \\
11131 & 3 & 5.8542 & 0.0611 & 0.010 \\
25874 & 3 & 4.9480 & 0.0081 & 0.002 \\
115383 & 15 & 6.2468 & 0.0523 & 0.008 \\
128620 & 4 & 5.4467 & 0.792 & 0.015 \\
138573 & 3 & 5.2340 & 0.1721 & 0.033 \\
146233 & 10 & 5.1494 & 0.0926 & 0.018 \\
147513 & 8 & 5.6597 & 0.0797 & 0.014 \\
165185 & 7 & 5.9908 & 0.1281 & 0.021 \\
182572 & 7 & 4.8009 & 0.0487 & 0.021 \\
190248 & 4 & 4.9663 & 0.0604 & 0.021 \\
196378 & 4 & 5.4779 & 0.0826 & 0.021 \\
206860 & 9 & 6.2659 & 0.0442 & 0.021 \\
221343 & 5 & 5.2848 & 0.0717 & 0.014 \\
\hline
  \end{tabular}
  \label{tab:repeatability}
\end{table}

\section{Results}
\label{sec5}

\subsection{H$\alpha$ chromospheric fluxes and \ion{Ca}{ii} H+K flux-flux relations}
Our large sample spanning wide intervals of T$_\text{eff}$, \feh and $\log{g}$, as well as mass, age, and activity levels, allows the detailed absolute flux-flux analysis between H$\alpha$ and \ion{Ca}{ii} ~H+K chromospheric losses, and its relation to intrinsic stellar properties established with high precision and consistency. \cite{lorenzo2016age} showed quantitatively that \ion{Ca}{ii} $\log(R^\prime_{HK})$ indexes are marred by strong mass and \feh biases, in the sense that metal-poor stars have shallower \ion{Ca}{ii} profiles that mimic high levels of chromospheric fill-in, and thus appear more active and younger than metal-rich stars at a given \teff or mass \citep[see also][]{rocha1998metallicity}. Such biases affect the derivation of age-activity relationships.

In order to explore these effects, we built a cross-sample for H$\alpha$ and \ion{Ca}{ii} H+K fluxes by compiling values of Mount Wilson $S_{MW}$ indexes from \cite{duncan1991ii,wright2004chromospheric,baliunas1995chromospheric,henry1996survey,gray2003contributions,gray2006contributions,jenkins2006activity,jenkins2008metallicities,mamajek2008improved,schroder2009ii,isaacson2010chromospheric,daSilva2021stellar}. We computed H+K chromospheric fluxes according to the prescription of \cite{middelkoop1982magnetic}. The photospheric correction $C_{cf}$, based on the $B-V$ colour, was used to determine the quantity $R_{HK}$, following the equation
\begin{equation}
    R_{HK} = 1.34 \times 10^{-4} \cdot C_{cf} \cdot S_{MW} \ .
\end{equation}

We applied the $R_{phot}$ correction of \cite{hartmann1984analysis}, which represents the residual photospheric contribution to the fluxes in these lines, and computed the $R'_{HK}$ quantity:
\begin{equation}
    R'_{HK} = R_{HK} - R_{phot} \ .
\end{equation}

We note that these calibrations, dating back to the creation of the Mount Wilson $<$S$>$ index, do not account for $[$Fe/H$]$ differences and are strictly valid only close to solar metallicity.

From the $R'_{HK}$ indexes, we arrived at the chromospheric losses in these lines ($F'_{HK}$), in erg cm$^{-2}$ s$^{-1}$, related to the bolometric fluxes by:
\begin{equation}
    \text{log}(R'_{HK}) = \text{log}\bigg(\frac{F'_{HK}}{\sigma \text{T}_\text{eff}^4}\bigg) \ .
\end{equation}

By checking the behavior of $F'_{HK}$ with \teff, we noticed a residual undesired trend for the hottest stars of the sample, which we interpreted as reminiscent photospheric fluxes insufficiently corrected for these stars. We corrected the trend by fitting an envelope to these fluxes, following
\begin{equation}
    \text{F}_{phot}^{HK} (\text{erg } \text{cm}^{-2} \text{ s}^{-1}) = - 8.48 \cdot \text{T}_\text{eff} + 2.65 \times 10^5 \ ,
\end{equation}
for stars with T$_\text{eff} < 5473$ K, and
\begin{equation}
    \text{F}_{phot}^{HK} (\text{erg } \text{cm}^{-2} \text{ s}^{-1}) = + 0.682 \cdot \text{T}_\text{eff}^2 -7428 \cdot \text{T}_\text{eff} + 2.04 \times 10^7 \ ,
\end{equation}
for T$_\text{eff} \geq 5473$ K. The subtraction of this envelope provides the final purely chromospheric losses in the H+K lines. This H$+$K flux scale leads to somewhat reduced fluxes and should not be taken at face value. It also reflects an arbitrary, sample-defined zero point, similar to our H$\alpha$ chromospheric flux scale, the important point being that both reflect real differences between active and inactive stars.

\begin{figure}
    \centering    \includegraphics[width=\columnwidth]{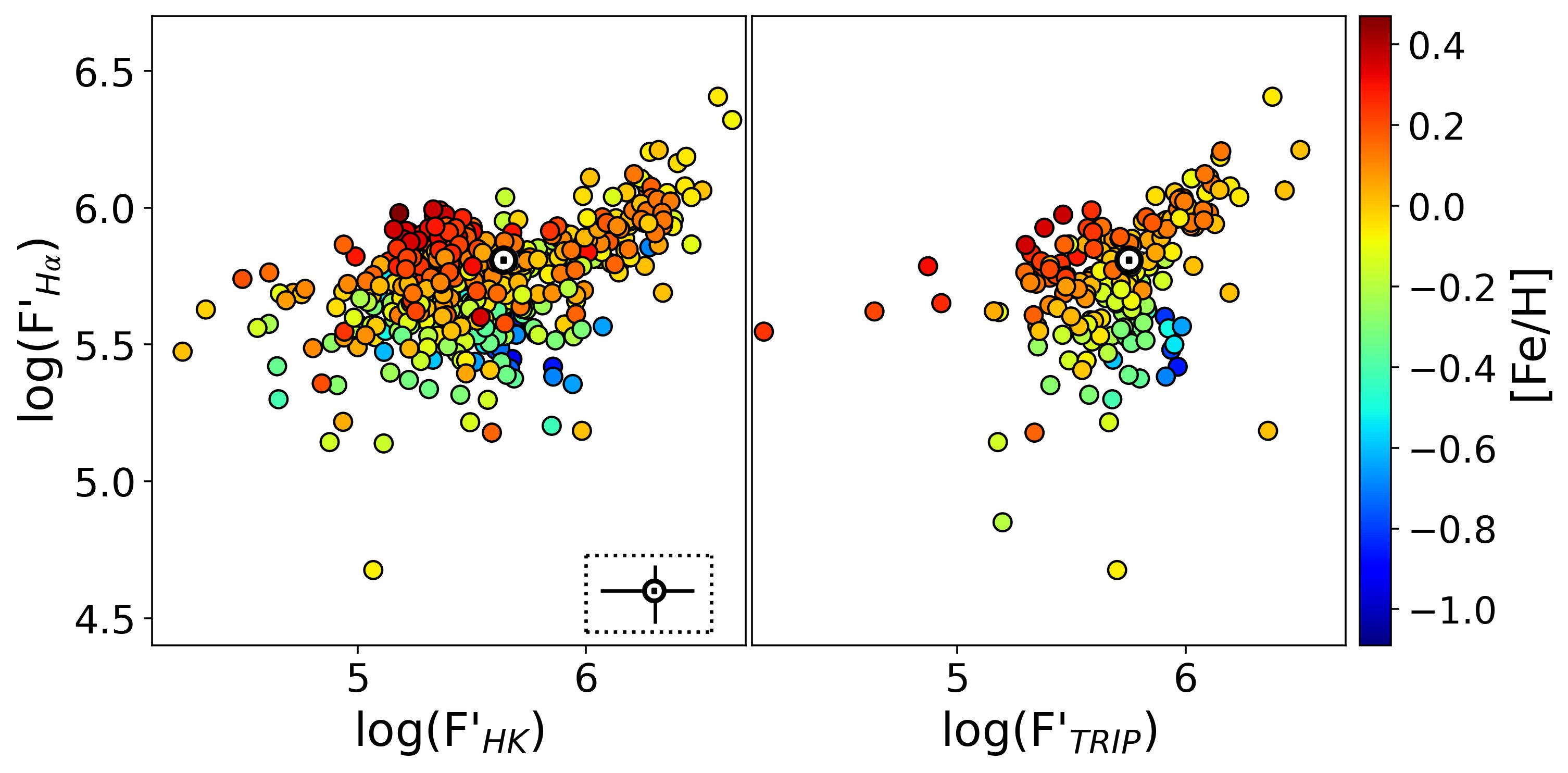}
    \caption{Relation between the H$\alpha$ chromospheric fluxes and $F'_{HK}$ (left panel) and $F'_{TRIP}$ (right panel), in erg $\text{cm}^{-2} \text{ s}^{-1}$, for stars with both measurements on each comparison. Typical error bars for the solar flux values are shown in the bottom right of the left panel. We assume the uncertainty in $F'_{HK}$ and $F'_{TRIP}$  to be $2 \times 10^5$ erg $\text{cm}^{-2} \text{ s}^{-1}$}.
    \label{fig:flux_flux}
\end{figure}

We next built the direct flux-flux correlation between the chromospheric fluxes of H$\alpha$ (here called $F'_{H\alpha}$) and the H+K lines ($F'_{HK}$), in erg $\text{cm}^{-2} \text{ s}^{-1}$, shown in Fig. \ref{fig:flux_flux} and color-coded by \feh. Not unexpectedly, the fluxes have a stronger correlation for the more active stars of our sample, composed chiefly of young stars of clusters and stellar associations. Below the $\sim10^{6}$ erg $\text{cm}^{-2} \text{ s}^{-1}$ level, for both variables, observational errors become important.

The most evident feature of the flux-flux comparison is that metal-rich stars occupy systematically higher positions than metal-poor stars, which means that their H$\alpha$/H+K flux ratio is consistently larger than the same ratio calculated for metal-poor stars. No comparable segregation is seen in similar \teff and $\log{g}$$-$stratified plots. This result is in line with the conclusions of \cite{lorenzo2016age} and is explained by purely observational biases; thereby, the deeper \ion{Ca}{ii} line profiles of metal-rich stars distort the true physical fluxes, making them appear reduced by lowering the photospheric correction baseline. In their turn, the shallower \ion{Ca}{ii} profiles artificially heighten the observed fluxes of metal-poor stars. This apparent distortion of H+K fluxes is seen in contrast to the H$\alpha$ fluxes because the Balmer line profiles are essentially insensitive to \feh variations \citep{fuhrmann1993balmer}. H$\alpha$ chromospheric losses can, therefore, be expected to better reflect actual physical chromospheric losses without a significant metallicity bias. This interpretation is independently reinforced by comparing $F'_{H\alpha}$ to $F'_{\rm TRIP}$ (right panel of Fig. \ref{fig:flux_flux}), the latter being the chromospheric fluxes in the \ion{Ca}{ii} infrared triplet lines derived by the prescription of Paper I.

\begin{figure}
    \centering
    \includegraphics[width=\columnwidth]{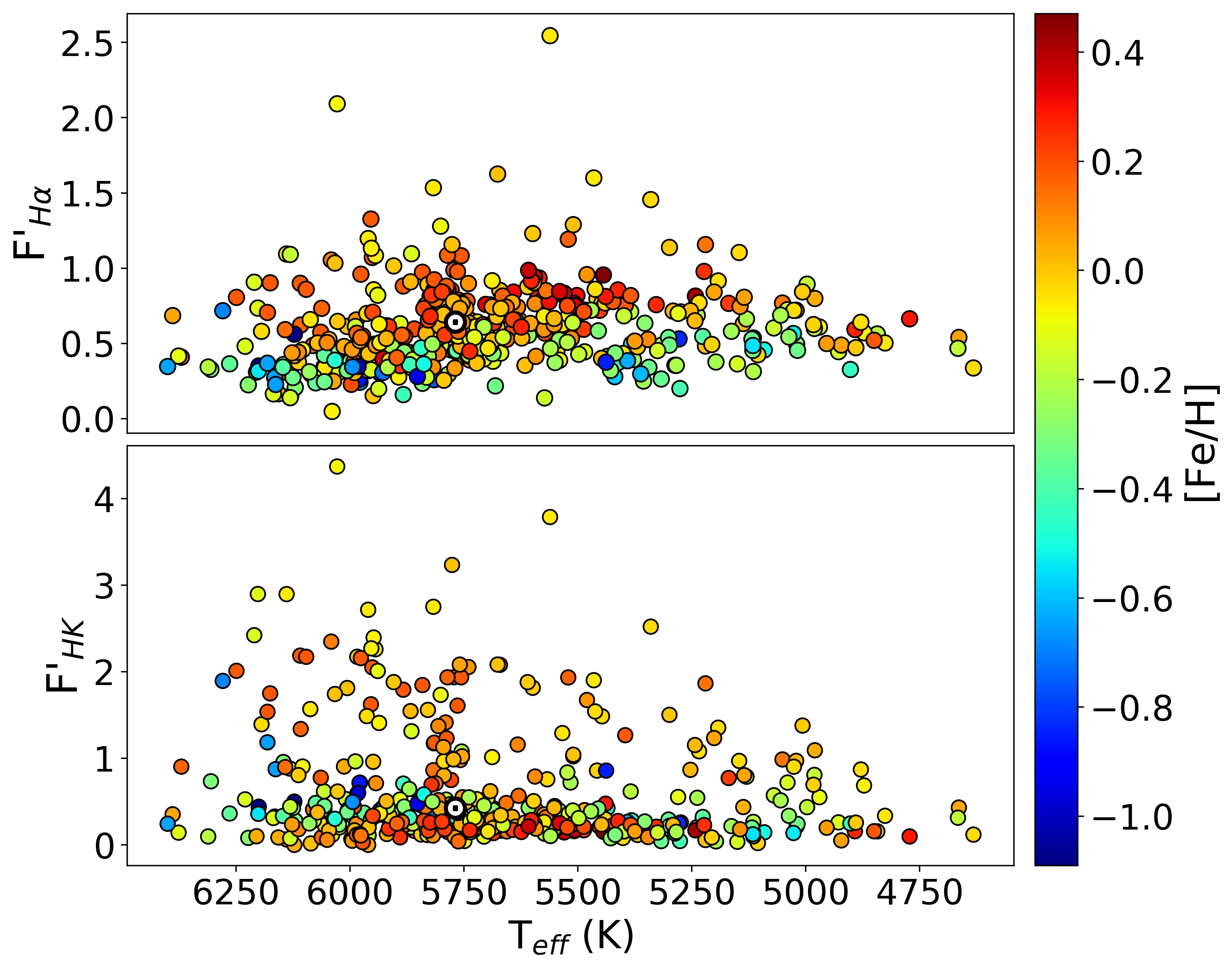}
    \caption{The run of $F'_{H\alpha}$ (top panel) and $F'_{HK}$ (bottom panel), in $10^6$ erg cm$^{\rm -2}$ s$^{\rm -1}$, {\it versus} \teff, color coded by $[$Fe/H$]$, for the 468 stars with both measurements. The Sun is plotted by its usual symbol.}
    \label{fig:FcromHa}
\end{figure}

We compare the chromospheric losses {\it versus} \teff for H$\alpha$ and the H+K lines (color-coded by [Fe/H]) in Fig. \ref{fig:FcromHa}. It is apparent that for H$\alpha$ the statistical expectation for the sample behavior is entirely borne out: metal-poor stars (older in average) populate lower values of $F'_{H\alpha}$, while metal-rich stars (younger in average) populate higher activity levels. This physically consistent behavior is absent in the H+K plot, in which metal-rich stars are found at very low levels of chromospheric activity. In contrast, metal-poor stars populate the high flux levels. Again, we interpret this odd state of affairs by invoking the aforementioned spectral biases, which affect chromospheric losses inferred from the \ion{Ca}{ii} metal lines but not H$\alpha$. 

To try and quantify the magnitude of the metallicity biases affecting $F'_{HK}$ as opposed to $F'_{H\alpha}$, we computed a linear regression to the data of Fig. \ref{fig:flux_flux}, and the residuals were fitted against [Fe/H]. Significant correlations were found: $\rho = 0.53$ and t$-$value $\sim$ 13.6 for the comparison between $F'_{H\alpha}$ and $F'_{HK}$, and $\rho = 0.60$ and t$-$value $\sim$ 10.8 for the comparison between $F'_{H\alpha}$ and $F'_{TRIP}$. At first-order, this behavior is well explained by the $[$Fe/H$]$ bias, and the remaining scatter is probably accounted for by variations in mass and evolutionary state, as well as cycle modulation.

\subsection{The age-activity relation}
The chromospheric activity {\it versus} age relation is built by retaining only stars with precise isochronal ages, here considered as $\sigma(\text{age}) \leq$ 1 Gyr. Members of open clusters and young stellar associations enter the relation by way of their collective age (Table \ref{tab:group_ages}). We deliberately excluded from the calibration the few stars with published asteroseismological ages and the Sun, which will be used later as a consistency check on the calibration built solely with isochronal ages.

\begin{table}
    \centering
    \caption{Stellar ages adopted for members of open clusters and young stellar associations.}
    \begin{tabular}{c|c|c}
    \hline
    \hline
         Cluster/Association  & Age & Reference \\
         \hline
         Beta Pictoris & $(25 \pm 3) \times 10^6$ yrs. & \cite{messina2016rotation} \\
         Hyades & $(760 \pm 35) \times 10^6$ yrs. & \cite{pasquini2023accurate} \\
         Pleiades & $(112 \pm 5) \times 10^6$ yrs. & \cite{dahm2015reexamining} \\
         Tucana-Horologium & $(45 \pm 4) \times 10^6$ yrs. & \cite{bell2015self} \\
    \hline
    \end{tabular}
    \label{tab:group_ages}
\end{table}

To build the calibration conservatively, we truncated the isochronal ages at $\leq$ 9 Gyr and the stellar F$'_{H\alpha}$ at $\geq$ 2 x 10$^{\rm 5}$ erg cm$^{\rm -2}$ s$^{\rm -1}$, taken as the lower limit of detectability. These cuts attempt to preclude ascribing unrealistic ages to stars with very low levels of chromospheric flux and to keep the ages physically within the age limit of the Galactic disc, taken at $\sim$9 Gyr \citep{del2005ageiii}, although we note that the distribution of absolute Gaia {\it G} magnitudes for white dwarfs within 40 pc of the Sun is compatible with older age, albeit significant uncertainties remain  \citep{cukanovaite2023local}. The final calibrating sample contains 186 stars $-$, a reduction of only 14 \% from the original 217 stars with precise isochronal ages. 

We performed the multilinear regression functionally using the same format as the calibration of \cite{lorenzo2016age}:
\begin{equation}
\begin{split}
    &\text{log} (\text{Age}) = \text{const.} + A \cdot \text{log} (F'_{H\alpha})+ B \cdot \text{log} (\text{M/M}_\odot) + C \cdot \text{[Fe/H]} \\ 
    &\ \ \ \ \ \ \ \ \ \ \ \ \ \  + D \cdot \text{log} (F'_{H\alpha})^2 , 
\end{split}
\label{eq:calib_age}
\end{equation}
for which the coefficients are given in Table \ref{tab:coefs_calib_age} with their errors and statistical t-values. Applying $|t| > 2$ criteria, all coefficients are relevant, bringing out the significant role played by mass and metallicity in setting the time evolution of chromospheric losses. We found R$^2 \approx 0.75$, and $\sigma_\text{reg} \approx 0.24$ dex (log(Age) in Gyr) as the standard error of the regression. This standard deviation can be contrasted to the relation found for the H+K age-activity relation by \cite{lorenzo2016age}, for which $\sigma_\text{reg} \approx 0.14$ dex.

We compare the ages determined by both our calibration (Eq. \ref{eq:calib_age}) and that of \cite{lorenzo2016age} for the \ion{Ca}{ii} H+K lines against published asteroseismological ages in Table \ref{tab:ages}. A simple linear regression with the latter as dependent variable provides correlation coefficients $\rho = 0.65$ and $\rho = 0.82$, respectively, for H$\alpha$ and \ion{Ca}{ii} H+K. Asteroseismological ages, though slightly model-dependent, are widely regarded as golden standards against which chromospheric ages must be measured. Stellar ages can be determined almost exclusively through model-dependent or empirical methods, no single method being accurately applicable across the full age range of the Galactic disc or for a wide range of masses \citep{soderblom2010ages}, therefore justifying an attack on the problem with different approaches. H$\alpha$ chromospheric losses suffer from larger relative errors than those derived from the H+K lines.

\begin{table}
    \centering
    \caption{Coefficients for the age-activity relation of Eq. \ref{eq:calib_age}, with their uncertainties and statistical t-values.}
    \begin{tabular}{c|c|c|c}
    \hline
    \hline
      Coefficient & Value & Error & $|t|$ \\
     \hline
     \rule{0pt}{2.5ex}
     constant & $-107.306$ & $11.753$ & 9.130 \\
     
     A & $+42.976$ & $4.047$ & 10.619\\
     
     B & $-4.279$ & $0.487$ & 8.785 \\
     
     C & $+0.835$  &  $0.152$ &  5.506  \\
     
     D & $-3.931$  &   0.348  & 11.281 \\
     
    \hline 
    \end{tabular}
    \label{tab:coefs_calib_age}
\end{table}

\begin{table}
    \centering
    \caption{Chromospheric ages derived from Eq. \ref{eq:calib_age} and published asteroseismological ages, in $10^9$ years, for the control stars. The first column is HD number, the second column and third columns are, respectively, the H$\alpha$ and \ion{Ca}{ii} chromospheric ages, last column provides the published asteroseismological age and its source: (a) \citet{huber202220}, (b) \citet{brandao2011asteroseismic}, (c) \citet{tang2011asteroseismic}, (d) \citet{ball2020robust}, (e) \citet{castro2021modeling}, (f) \citet{chontos2021tess}, (g) \citet{joyce2018classically}, (h) \citet{metcalfe2023asteroseismology}, (i) \citet{Bazot_2018}, (j) \citet{soriano2010new}, (k) \citet{mosser2008hd}, (l) \citet{metcalfe2024weakened}, (m) \citet{metcalfe2020evolution} and (n) \citet{connelly2012absolute}.}
    \begin{tabular}{c|c|c|c|c}
    \hline
    \hline
        Star & H$\alpha$ Age & H+K Age & Aster. Age \\
        \hline
        1581 & $8.15 \substack{+6.01 \\ -3.46}$ & $5.79 \substack{+2.20 \\ -1.60}$ & $5.30 \pm 0.50$ (a) \\[0.1cm]
        2151 & $7.25 \substack{+5.35 \\ -3.07}$ & $5.73 \substack{+2.18 \\ -1.58}$ & $6.40 \pm 0.56$ (b) \\[0.1cm]
        10700 & $10.83 \substack{+7.99 \\ -4.60}$ & $14.07 \substack{+5.35 \\ -3.88}$ & [8 - 10] (c) \\[0.1cm]
        38529 & $2.00 \substack{+1.48 \\ -0.85}$ & $2.71 \substack{+1.03 \\ -0.75}$ & $3.07 \pm 0.39$ (d) \\[0.1cm]
        43587 & $11.03 \substack{+8.14 \\ -4.68}$ & $6.16 \substack{+2.34 \\ -1.70}$ & $6.2 \pm 0.1$ (e) \\[0.1cm]
        43834 & $8.64 \substack{+6.38 \\ -3.67}$ & $5.15 \substack{+1.96 \\ -1.42}$ & $6.2 \pm 1.4$ (f) \\[0.1cm]
        128620 & $2.49 \substack{+1.83 \\ -1.05}$ & $4.46 \substack{+1.70 \\ -1.23}$ & $5.3 \pm 0.3$ (g) \\[0.1cm]
        128621 & $3.02 \substack{+2.25 \\ -1.28}$ & $5.23 \substack{+1.99 \\ -1.44}$ & $5.3 \pm 0.3$ (g) \\[0.1cm]
        141004 & $9.57 \substack{+7.06 \\ -4.06}$ & $5.62 \substack{+2.14 \\ -1.55}$ & $5.40 \pm 0.70$ (h) \\[0.1cm]
        146233 & $5.70 \substack{+4.20 \\ -2.42}$ & $5.36 \substack{+2.04 \\ -1.48}$ & $4.67  \substack{+0.87 \\ -1.29}$ (i) \\[0.1cm]
        160691 & $5.96 \substack{+4.39 \\ -2.53}$ & $4.55 \substack{+1.73 \\ -1.25}$ & $6.34 \pm 0.80$ (j) \\[0.1cm]
        203608 & $6.24 \substack{+4.60 \\ -2.65}$ & $4.93 \substack{+1.88 \\ -1.36}$ & $7.25 \pm 0.07$ (k) \\[0.1cm]
        217014 & $2.37 \substack{+1.74 \\ -1.00}$ & $4.59 \substack{+1.75 \\ -1.26}$ & $4.80 \substack{+0.70 \\ -0.40}$ (l) & \\[0.1cm]
        219834 & $4.13 \substack{+3.05 \\ -1.76}$ & $4.64 \substack{+1.76 \\ -1.28}$ & $6.20 \pm 0.20$ (m) & \\[0.1cm]
        Sun & $5.06 \substack{+3.73 \\ -2.15}$ & $5.29 \substack{+2.01 \\ -1.46}$ & $4.5670 \pm 0.0002$ (n) & \\
    \hline
    \end{tabular}
    \label{tab:ages}
\end{table}

The calibration of Eq. \ref{eq:calib_age} expresses the different timescales of the decay of chromospheric activity obtained for stars with distinct mass and $[$Fe/H$]$, and can be contrasted to the age-activity calibration of the Eq. 1 of \citet{lorenzo2016age},
\begin{equation}
\begin{split}
    \text{log}(t) = -56.01 - 25.81 \cdot \text{log} (R'_{HK}) - 0.44 \cdot \text{[Fe/H]} \\
    - 1.26 \cdot \text{log} (\text{M/M}_\odot) - 2.53 \cdot \text{log} (R'_{HK})^2 \ .
\end{split}
\label{eq:calib_diego}
\end{equation}
The theoretical expectation to be checked with these two calibrations is based on the surface convection of solar-type stars and on the chromospheric activity decay over time:
\begin{itemize}
    \item For two stars with the same [Fe/H] and the same level of chromospheric activity, the more massive star, convectively less efficient, needs to be younger than the less massive star, convectively more efficient, to explain the observed activity;
    \item For two stars with the same mass and level of chromospheric activity, the metal$-$poorer star, convectively less efficient, needs to be younger than the richest star, convectively more efficient, to explain the observed activity.
\end{itemize}
The first point is verified in both calibrations through the negative sign of the mass coefficient. Reducing mass increases log(\textit{t}), and vice-versa. The second point is observed only in our H$\alpha$ calibration, where the metallicity coefficient is positive. Increasing [Fe/H] would increase log(\textit{t}), and vice-versa. As we proposed, the apparent failure of the H+K calibration in recovering this feature is due to the metallicity bias discussed in the last subsection. From a purely theoretical point of view, mass and metallicity are probably not fully independent as variables, but the qualitative behavior attached to our calibration seems robust and in line with theoretical expectation, even though we do not support a strict quantitative determination owing to the intrinsically large uncertainties associated with the H$\alpha$ chromospheric fluxes, the isochronal ages and the regressive model itself, which is certainly a poor representation of an intrinsically complex problem.

\begin{figure}
    \centering
    \includegraphics[width=6.9cm]{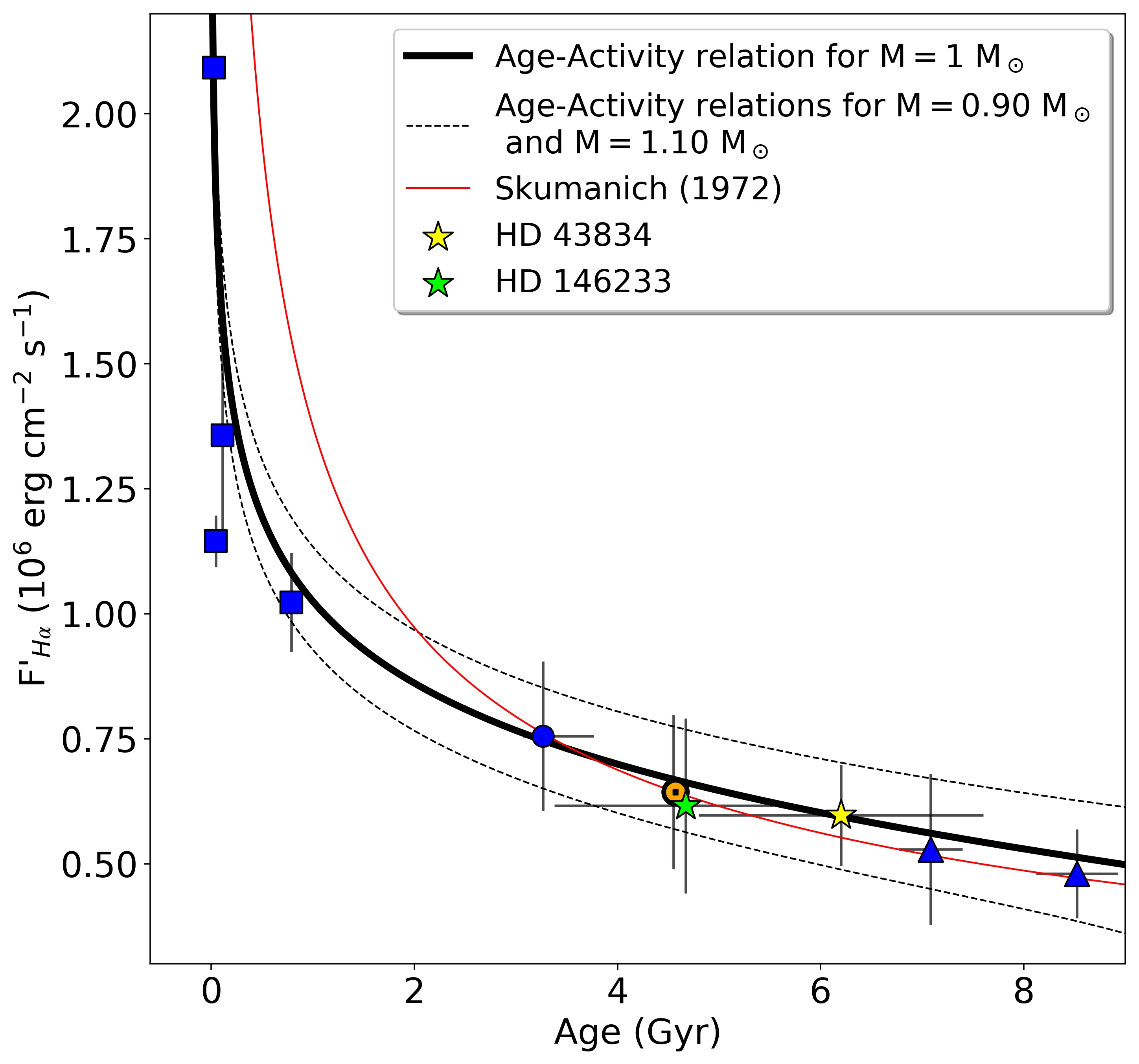}
    \caption{The age activity-relation from Eq. \ref{eq:calib_age} for 1.00 M$_\odot$ and solar metallicity (black line), with data points in the range 0.90 $<$ M/M$_\odot$ $<$ 1.10 and $-0.20$ $<$ $[$Fe/H$]$ $<$ $+$0.20, sided by the corresponding relations for 0.90 M/M$_\odot$ (upper dashed line) and 1.10 M/M$_\odot$ (lower dashed line). The Sun is plotted with its usual symbol (orange) fitted by a Skumanich-like decay law (red line).}
    \label{fig:age_activity_solartype}
\end{figure}

\begin{figure}
    \centering
    \includegraphics[width=7.8cm]{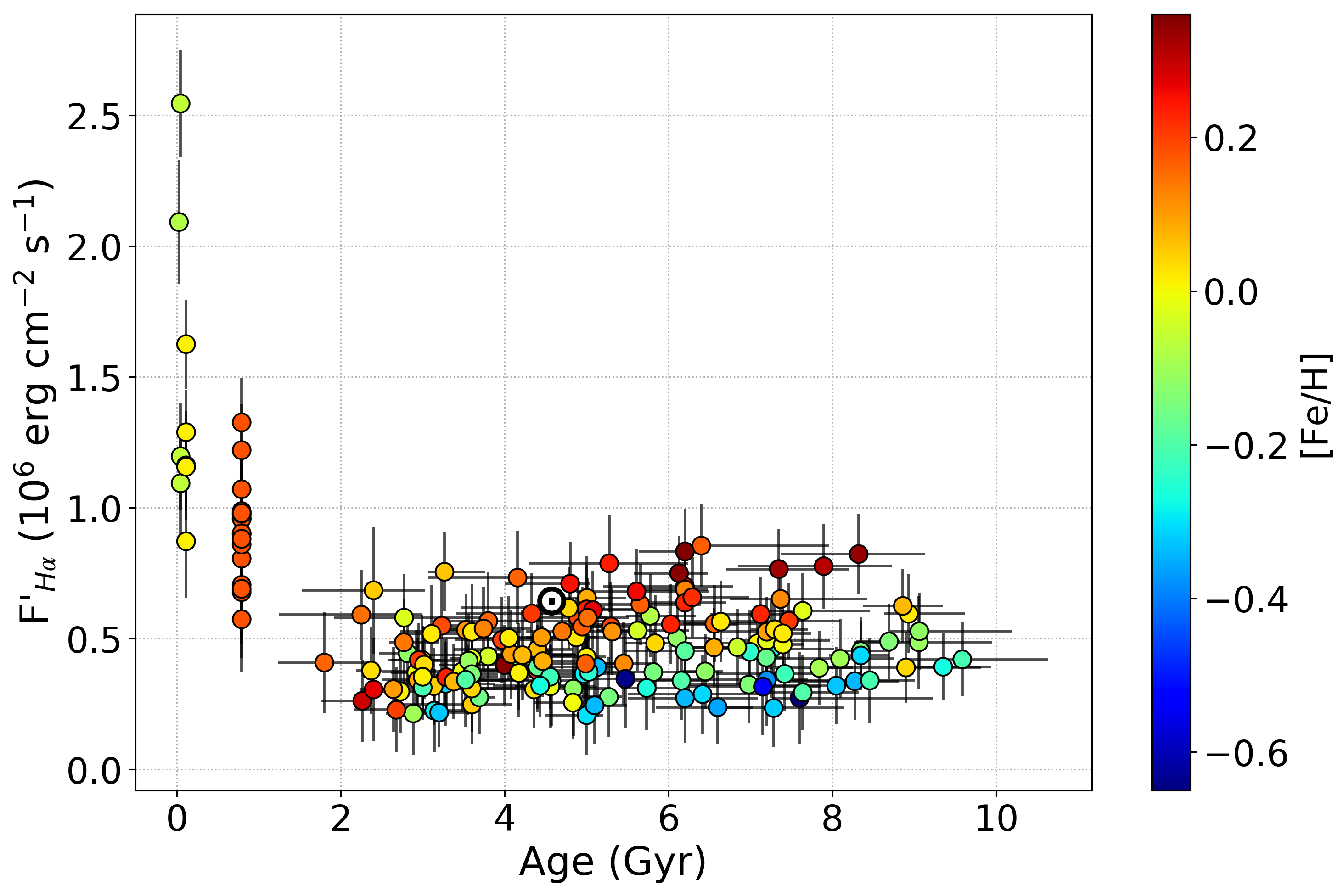}
    \caption{Run of H$\alpha$ chromospheric fluxes with the stellar isochronal ages (color-coded by $[$Fe/H$]$) for the 186 stars that define the regression of Eq. \ref{eq:calib_age} (plus the Sun), namely, all stars for which $\sigma_{\rm age}$ $<$ 1 Gyr, $F'_{H\alpha}$ values $>$ 2 $\times$ 10$^{\rm 5}$ erg cm$^{\rm -2}$ s$^{\rm -1}$ and isochronal ages $<$ 9 Gyr (see text). The Sun is plotted with its usual symbol.}
    \label{fig:age_activity_general}
\end{figure}

The age-activity relation is superimposed to the data points of stars in a narrow range of \feh and mass in Fig. \ref{fig:age_activity_solartype}, including the Pleiades and Hyades clusters, the $\beta$ Pic and Tuc-Hor young associations, and two stars with asteroseismological ages. We averaged 15 stars with isochronal ages in the 6.0$-$7.5 Gyr and 7.5$-$9.0 Gyr age ranges into two data points, for which the uncertainties are the averages of the age and chromospheric flux errors of the individual objects. By restricting the mass and metallicity dimensions, we found an age-activity trend very similar to the one found by \cite{lorenzo2018solar} for solar twin stars. We also note that a Skumanich-like power law (proportional to age$^{-0.5}$), forced to pass through the solar position, initially underestimates the rapidity of the decay in chromospheric flux but agrees with our calibration at $\sim$3 Gyr. In Fig. \ref{fig:age_activity_general}, the \feh and mass dimensions are ignored, and the age-activity relation is all but lost $-$ very similarly to Fig.1 of \cite{pace2013chromospheric}. One also notes that similarly to Fig. \ref{fig:FcromHa}, metal-rich stars (younger on average) populate higher levels of activity than metal-poor stars (older on average). As a further test, we isolated a subsample of Fig. \ref{fig:age_activity_general} in the narrow 0.90 $<$ M/M{$_\odot$} $<$ 1.10 range and ascertained that the aforementioned effect is even more clearly expressed.

H$\alpha$ chromospheric ages for our sample of 511 stars are given in Table \ref{tab:chrom_ages}, along with chromospheric ages determined through the calibration of \cite{lorenzo2016age} (Eq. \ref{eq:calib_diego}, their Eq. 1), using $R'_{HK}$  (whenever available). Conservatively, we did not attempt to derive H$\alpha$ chromospheric ages for stars with $F'_{H\alpha}$ $\leq 2\times 10^5$ erg cm$^{-2}$ s$^{-1}$, as explained above. Similarly, we refrain from ascribing ages to stars with $R'_{HK}$ values below $-$5.1: this value corresponds to extremely inactive \& evolved stars \citep{wright2004}. Moreover, we reject chromospheric age determinations that surpass $\geq$9 Gyr. Stars falling into one or more of the criteria above are included in Table \ref{tab:chrom_ages} but have empty entries.

\begin{table}
    \centering
    \caption{Chromospheric ages, in $10^9$ years, for the sample stars (identified in the first column), calculated by the calibration of Eq. \ref{eq:calib_age} (second column) from the H$\alpha$ chromospheric fluxes and the calibration of \citet{lorenzo2016age} (third column) from the H+K chromospheric fluxes (when available). See text for details. The full table is available online.}
    \begin{tabular}{c|c|c}
    \hline
    \hline
    Star & H$\alpha$ Age & H+K Age 
    \\ & (this work) & \citep{lorenzo2016age} \\
    \hline
HD105 & 0.379$\substack{+0.280 \\- 0.161} $ & 0.274$\substack{+0.104 \\- 0.076} $ \\[0.1cm]
HD166 & 0.822$\substack{+0.606 \\- 0.349} $ & 0.299$\substack{+0.114 \\- 0.082} $ \\[0.1cm]
HD1237 & 2.087$\substack{+1.540 \\- 0.886} $ & 0.486$\substack{+0.185 \\- 0.134} $ \\[0.1cm]
HD1461 & 3.248$\substack{+2.396 \\- 1.379} $ & 4.968$\substack{+1.890 \\- 1.369} $ \\[0.1cm]
HD1466 & 0.447$\substack{+0.330 \\- 0.190} $ & 0.273$\substack{+0.104 \\- 0.075} $ \\[0.1cm]
HD1581 & 8.152$\substack{+6.014 \\- 3.461} $ & 5.792$\substack{+2.203 \\- 1.596} $ \\[0.1cm]
HD1835 & 1.255$\substack{+0.926 \\- 0.533} $ & 0.429$\substack{+0.163 \\- 0.118} $ \\[0.1cm]
HD2151 & 7.252$\substack{+5.350 \\- 3.079} $ & 5.727$\substack{+2.179 \\- 1.578} $ \\[0.1cm]
HD3047 & 3.680$\substack{+2.715 \\- 1.562} $ & 4.899$\substack{+1.863 \\- 1.350} $ \\[0.1cm]
HD3443 & 6.833$\substack{+5.041 \\- 2.901} $ & 7.193$\substack{+2.736 \\- 1.982} $ \\
... & ... & ... \\
    \hline
    \end{tabular}
    \label{tab:chrom_ages}
\end{table}

\section{Conclusions}
\label{sec6}
We determined the absolute H$\alpha$ total and purely chromospheric fluxes for 511 solar-type stars spanning a wide interval of precisely determined masses, metallicities, ages, and states of evolution through the use of modern model atmospheres and spectra of high S/N ratio and moderately high resolution. A careful determination of the stellar parameters was made through photometric calibrations of effective temperature, the compilation of spectroscopic determinations of \feh from the literature, the use of recent evolutionary tracks and isochrones to estimate mass and age, and the HIPPARCOS and Gaia DR2 parameters as tools for the derivation of the remaining variables. We also compiled for the sample stars published values of $S_{MW}$ and computed their H+K chromospheric losses. Both $F'_{H\alpha}$ and $F'_{HK}$ are expressed in a physical basis as erg $\text{cm}^{-2} \text{ s}^{-1}$. We discussed in detail the flux-flux relation for H$\alpha$ and the H+K lines and quantified the \feh biases that strongly affect the physical interpretation of the \ion{Ca}{ii} H+K fluxes. We derive for the H$\alpha$ fluxes a detailed age-activity-mass-metallicity calibration for 186 stars with precise isochronal ages.

We summarize our main conclusions as follows: 
\begin{enumerate}
    \item Even though total H$\alpha$ chromospheric losses are of a lesser magnitude and less precise than those of the \ion{Ca}{ii} H+K lines, they are useful diagnostics of magnetic activity, spanning a sizable dynamic range and well correlated both to $F'_{HK}$ and \ion{Ca}{ii} infrared triplet fluxes, except for the lowest activity levels;
    
    \item The underlying photospheric flux of the \ion{Ca}{ii} lines, as well as that of the \ion{Ca}{ii} infrared triplet lines, suffers from a strong bias in which the deep profiles of metal-rich stars (in average, younger) mimic lower chromospheric fluxes, while the shallow profiles of metal-poor stars (in average, older) mimic enhanced chromospheric losses, significantly blurring the age-activity relation. This bias accounts for most (if not all) of the absence of chromospheric activity decay with age after $\sim$1 Gyr, reported in the literature;

    \item The \ion{Ca}{ii} H+K lines being the standard, most widely studied tool to quantify magnetic activity in solar-type stars, care should be exercised in employing them whenever there is reason to suppose that wide ranges of mass and/or \feh may be involved;
    
    \item H$\alpha$ chromospheric losses do not suffer from significant metallicity biases and, despite suffering from larger uncertainties in chromospheric flux, are shown to be useful age indicators for FGK stars, particularly if employed in tandem with other age diagnostics;
    
    \item The H$\alpha$'s age-activity-mass-metallicity calibration appears to be in line with the theoretical expectation that (all other parameters being equal) more massive stars possess narrower convective zones and are less active than less massive stars. In contrast, more metal-rich stars, with their deeper convective zones, appear more active than metal-poorer stars.
\end{enumerate}

Clear directions of improvement on our analysis is to build a larger sample, providing a better understanding of the metallicity bias affecting the \ion{Ca}{ii} lines and a more precise estimation of the zero-point flux scale of H$\alpha$. Our work supports the notion that chromospheric fluxes allow the determination of meaningful ages up to the solar age, particularly if multiplex chromospheric age determinations can be brought to bear simultaneously. Even though H$\alpha$ chromospheric losses suffer from significant relative errors, we suggest it is sufficiently precise to add statistical significance to chromospheric ages when averaged between various spectroscopic indicators and completely independent methods. Such compositions of independent methods to constrain stellar ages are gaining traction, the most widely used, besides chromospheric activity, being Li abundances \citep{mamajek2002post,nielsen2010uniform,jeffries2023gaia}, kinematics \citep{almeida2018method}, gyrochronology \citep[][who also argue in favor of chromospheric methods to uncover undetected companions]{barnes2007ages} and chemical clocks \citep{da2012accurate}, to cite a few. Suitable recent applications of such a syncretic approach are given by \cite{burgasser2017age} for the TRAPPIST-1 red dwarf system, whereby the authors employ Li abundances, surface gravity features, metallicity, kinematics, rotation, and magnetic activity, among other criteria, to constrain the system's age at $\sim$7.6 Gyr; and by \cite{stanford2020baffles}, employing colours, Li absorption strengths and R'$_{\rm HK}$ indices to obtain posterior ages for 2630 nearby field stars.

\section*{Acknowledgements}

This paper is based on the senior monograph of P.V.S.S., who also acknowledges a CAPES/Brazil PhD scholarship under grant 88887.821758/2023-00; E.C.B. acknowledges a CNPq/Brazil scholarship; G.F.P.M. acknowledges financial support from CNPq/Brazil under grant 474972/2009-7. D.L.-O. thanks CNPq grant PCI$-$301612/2024-2. F.A.-F. acknowledges funding from FAPESP/Brazil grant 2018/20977-2. I.R. acknowledges financial support from the Agencia Estatal de Investigaci\'on of the Ministerio de Ciencia e Innovaci\'on MCIN/AEI/10.13039/501100011033 and the ERDF ``A way of making Europe'' through project PID2021-125627OB-C31, from the Centre of Excellence ``Mar\'{\i}a de Maeztu'' award to the Institut de Ci\`encies de l’Espai (CEX2020-001058-M) and from the Generalitat de Catalunya/CERCA programme. We thank the staff of OPD/LNA for their considerable support during the many observation runs carried out during this project. Use was made of the Simbad database, operated at the CDS, Strasbourg, France, and of NASA’s Astrophysics Data System Bibliographic Services. This work presents results from the European Space Agency (ESA) space mission Gaia. We thank the anonymous referee for criticism and suggestions that considerably improved this paper.

\section*{Data Availability}

Full tables 3, 4, 5 and 10 are available online.



\bibliographystyle{mnras}
\bibliography{bibliography} 

\begin{thebibliography}{}
\makeatletter
\relax
\def\mn@urlcharsother{\let\do\@makeother \do\$\do\&\do\#\do\^\do\_\do\%\do\~}
\def\mn@doi{\begingroup\mn@urlcharsother \@ifnextchar [ {\mn@doi@} {\mn@doi@[]}}
\def\mn@doi@[#1]#2{\def\@tempa{#1}\ifx\@tempa\@empty \href {http://dx.doi.org/#2} {doi:#2}\else \href {http://dx.doi.org/#2} {#1}\fi \endgroup}
\def\mn@eprint#1#2{\mn@eprint@#1:#2::\@nil}
\def\mn@eprint@arXiv#1{\href {http://arxiv.org/abs/#1} {{\tt arXiv:#1}}}
\def\mn@eprint@dblp#1{\href {http://dblp.uni-trier.de/rec/bibtex/#1.xml} {dblp:#1}}
\def\mn@eprint@#1:#2:#3:#4\@nil{\def\@tempa {#1}\def\@tempb {#2}\def\@tempc {#3}\ifx \@tempc \@empty \let \@tempc \@tempb \let \@tempb \@tempa \fi \ifx \@tempb \@empty \def\@tempb {arXiv}\fi \@ifundefined {mn@eprint@\@tempb}{\@tempb:\@tempc}{\expandafter \expandafter \csname mn@eprint@\@tempb\endcsname \expandafter{\@tempc}}}

\bibitem[\protect\citeauthoryear{Adibekyan, Sousa, Santos, Mena, Hern{\'a}ndez, Israelian, Mayor  \& Khachatryan}{Adibekyan et~al.}{2012}]{adibekyan2012chemical}
Adibekyan V.~Z.,  Sousa S.,  Santos N.,  Mena E.~D.,  Hern{\'a}ndez J.~G.,  Israelian G.,  Mayor M.,   Khachatryan G.,  2012, \mn@doi [Astronomy \& Astrophysics] {10.1051/0004-6361/201219401}, 545, A32

\bibitem[\protect\citeauthoryear{Aguilera-G{\'o}mez, Ram{\'\i}rez  \& Chanam{\'e}}{Aguilera-G{\'o}mez et~al.}{2018}]{aguilera2018lithium}
Aguilera-G{\'o}mez C.,  Ram{\'\i}rez I.,   Chanam{\'e} J.,  2018, \mn@doi [Astronomy \& Astrophysics] {10.1051/0004-6361/201732209}, 614, A55

\bibitem[\protect\citeauthoryear{Almeida-Fernandes \& Rocha-Pinto}{Almeida-Fernandes \& Rocha-Pinto}{2018}]{almeida2018method}
Almeida-Fernandes F.,  Rocha-Pinto H.~J.,  2018, Monthly Notices of the Royal Astronomical Society, 476, 184

\bibitem[\protect\citeauthoryear{Almeida-Fernandes et~al.,}{Almeida-Fernandes et~al.}{2023}]{almeida2023chemodynamical}
Almeida-Fernandes F.,  et~al., 2023, \mn@doi [Monthly Notices of the Royal Astronomical Society] {10.1093/mnras/stad1561}, 523, 2934

\bibitem[\protect\citeauthoryear{Almeida, Santos, Melo, Ammler-von Eiff, Torres, Quast, Gameiro  \& Sterzik}{Almeida et~al.}{2009}]{almeida2009search}
Almeida P.~V.,  Santos N.,  Melo C.,  Ammler-von Eiff M.,  Torres C.,  Quast G.,  Gameiro J.,   Sterzik M.,  2009, \mn@doi [Astronomy \& Astrophysics] {10.1051/0004-6361/200811194}, 501, 965

\bibitem[\protect\citeauthoryear{Amard \& Matt}{Amard \& Matt}{2020}]{amard2020impact}
Amard L.,  Matt S.~P.,  2020, \mn@doi [The Astrophysical Journal] {10.3847/1538-4357/ab6173}, 889, 108

\bibitem[\protect\citeauthoryear{Baliunas et~al.,}{Baliunas et~al.}{1995}]{baliunas1995chromospheric}
Baliunas S.,  et~al., 1995, \mn@doi [Astrophysical Journal, Part 1 (ISSN 0004-637X), vol. 438, no. 1, p. 269-287] {10.1086/175072}, 438, 269

\bibitem[\protect\citeauthoryear{Ball et~al.,}{Ball et~al.}{2020}]{ball2020robust}
Ball W.~H.,  et~al., 2020, \mn@doi [Monthly Notices of the Royal Astronomical Society] {10.1093/mnras/staa3190}, 499, 6084

\bibitem[\protect\citeauthoryear{Barnes}{Barnes}{2007}]{barnes2007ages}
Barnes S.~A.,  2007, \mn@doi [The Astrophysical Journal] {10.1086/519295}, 669, 1167

\bibitem[\protect\citeauthoryear{Barnes \& Kim}{Barnes \& Kim}{2010}]{barnes2010angular}
Barnes S.~A.,  Kim Y.-C.,  2010, \mn@doi [The Astrophysical Journal] {10.1088/0004-637X/721/1/675}, 721, 675

\bibitem[\protect\citeauthoryear{Barry}{Barry}{1988}]{barry1988chromospheric}
Barry D.~C.,  1988, \mn@doi [The Astrophysical Journal] {10.1086/166848}, 334, 436

\bibitem[\protect\citeauthoryear{Barry, Cromwell  \& Hege}{Barry et~al.}{1987}]{barry1987chromospheric}
Barry D.~C.,  Cromwell R.~H.,   Hege E.~K.,  1987, \mn@doi [The Astrophysical Journal] {10.1086/165131}, 315, 264

\bibitem[\protect\citeauthoryear{Bazot, Creevey, Christensen-Dalsgaard  \& Meléndez}{Bazot et~al.}{2018}]{Bazot_2018}
Bazot M.,  Creevey O.,  Christensen-Dalsgaard J.,   Meléndez J.,  2018, \mn@doi [Astronomy and Astrophysics] {10.1051/0004-6361/201834058}, 619, A172

\bibitem[\protect\citeauthoryear{Bell, Mamajek  \& Naylor}{Bell et~al.}{2015}]{bell2015self}
Bell C.~P.,  Mamajek E.~E.,   Naylor T.,  2015, \mn@doi [Monthly Notices of the Royal Astronomical Society] {10.1093/mnras/stv1981}, 454, 593

\bibitem[\protect\citeauthoryear{Bensby, Feltzing  \& Oey}{Bensby et~al.}{2014}]{bensby2014exploring}
Bensby T.,  Feltzing S.,   Oey M.,  2014, \mn@doi [Astronomy \& Astrophysics] {10.1051/0004-6361/201322631}, 562, A71

\bibitem[\protect\citeauthoryear{Boesgaard \& Friel}{Boesgaard \& Friel}{1990}]{boesgaard1990chemical}
Boesgaard A.~M.,  Friel E.~D.,  1990, \mn@doi [The Astrophysical Journal] {10.1086/168484}, 351, 467

\bibitem[\protect\citeauthoryear{Boesgaard, Armengaud, King, Deliyannis  \& Stephens}{Boesgaard et~al.}{2004}]{boesgaard2004correlation}
Boesgaard A.~M.,  Armengaud E.,  King J.~R.,  Deliyannis C.~P.,   Stephens A.,  2004, \mn@doi [The Astrophysical Journal] {10.1086/423194}, 613, 1202

\bibitem[\protect\citeauthoryear{Brand{\~a}o et~al.,}{Brand{\~a}o et~al.}{2011}]{brandao2011asteroseismic}
Brand{\~a}o I.,  et~al., 2011, \mn@doi [Astronomy \& Astrophysics] {10.1051/0004-6361/201015370}, 527, A37

\bibitem[\protect\citeauthoryear{Bressan, Marigo, Girardi, Salasnich, Dal~Cero, Rubele  \& Nanni}{Bressan et~al.}{2012}]{bressan2012parsec}
Bressan A.,  Marigo P.,  Girardi L.,  Salasnich B.,  Dal~Cero C.,  Rubele S.,   Nanni A.,  2012, \mn@doi [Monthly Notices of the Royal Astronomical Society] {10.1111/j.1365-2966.2012.21948.x}, 427, 127

\bibitem[\protect\citeauthoryear{Buder et~al.,}{Buder et~al.}{2018}]{buder2018galah}
Buder S.,  et~al., 2018, \mn@doi [Monthly Notices of the Royal Astronomical Society] {10.1093/mnras/sty1281}, 478, 4513

\bibitem[\protect\citeauthoryear{Buder et~al.,}{Buder et~al.}{2019}]{buder2019galah}
Buder S.,  et~al., 2019, \mn@doi [Astronomy \& Astrophysics] {10.1051/0004-6361/201833218}, 624, A19

\bibitem[\protect\citeauthoryear{Burgasser \& Mamajek}{Burgasser \& Mamajek}{2017}]{burgasser2017age}
Burgasser A.~J.,  Mamajek E.~E.,  2017, The Astrophysical Journal, 845, 110

\bibitem[\protect\citeauthoryear{Bus{\`a}, Cuadrado, Terranegra, Andretta  \& Gomez}{Bus{\`a} et~al.}{2007}]{busa2007ii}
Bus{\`a} I.,  Cuadrado R.~A.,  Terranegra L.,  Andretta V.,   Gomez M.,  2007, \mn@doi [Astronomy \& Astrophysics] {10.1051/0004-6361:20065588}, 466, 1089

\bibitem[\protect\citeauthoryear{Carrera et~al.,}{Carrera et~al.}{2019}]{carrera2019open}
Carrera R.,  et~al., 2019, \mn@doi [Astronomy \& Astrophysics] {10.1051/0004-6361/201834546}, 623, A80

\bibitem[\protect\citeauthoryear{Casagrande, Ram{\'\i}rez, Melendez, Bessell  \& Asplund}{Casagrande et~al.}{2010}]{casagrande2010absolutely}
Casagrande L.,  Ram{\'\i}rez I.,  Melendez J.,  Bessell M.,   Asplund M.,  2010, \mn@doi [Astronomy \& Astrophysics] {10.1051/0004-6361/200913204}, 512, A54

\bibitem[\protect\citeauthoryear{Casagrande et~al.,}{Casagrande et~al.}{2014}]{casagrande2014towards}
Casagrande L.,  et~al., 2014, \mn@doi [Monthly Notices of the Royal Astronomical Society] {10.1093/mnras/stu089}, 439, 2060

\bibitem[\protect\citeauthoryear{Casagrande et~al.,}{Casagrande et~al.}{2021}]{casagrande2021galah}
Casagrande L.,  et~al., 2021, \mn@doi [Monthly Notices of the Royal Astronomical Society] {10.1093/mnras/stab2304}, 507, 2684

\bibitem[\protect\citeauthoryear{Castro et~al.,}{Castro et~al.}{2021}]{castro2021modeling}
Castro M.,  et~al., 2021, \mn@doi [Monthly Notices of the Royal Astronomical Society] {10.1093/mnras/stab1410}, 505, 2151

\bibitem[\protect\citeauthoryear{Cayrel~de Strobel \& Bentolila}{Cayrel~de Strobel \& Bentolila}{1989}]{cayrel1989search}
Cayrel~de Strobel G.,  Bentolila C.,  1989, Astronomy and Astrophysics, 211, 324

\bibitem[\protect\citeauthoryear{Chaffee~Jr, Carbon  \& Strom}{Chaffee~Jr et~al.}{1971}]{chaffee1971abundances}
Chaffee~Jr F.~H.,  Carbon D.~F.,   Strom S.~E.,  1971, \mn@doi [The Astrophysical Journal] {10.1086/150985}, 166, 593

\bibitem[\protect\citeauthoryear{Chmielewski}{Chmielewski}{2000}]{chmielewski2000infrared}
Chmielewski Y.,  2000, Astronomy and Astrophysics, v. 353, p. 666-690 (2000), 353, 666

\bibitem[\protect\citeauthoryear{Chontos et~al.,}{Chontos et~al.}{2021}]{chontos2021tess}
Chontos A.,  et~al., 2021, \mn@doi [The Astrophysical Journal] {10.3847/1538-4357/ac1269}, 922, 229

\bibitem[\protect\citeauthoryear{Clegg, Lambert  \& Tomkin}{Clegg et~al.}{1981}]{clegg1981carbon}
Clegg R.,  Lambert D.,   Tomkin J.,  1981, \mn@doi [The Astrophysical Journal] {10.1086/159371}, 250, 262

\bibitem[\protect\citeauthoryear{Connelly, Bizzarro, Krot, Nordlund, Wielandt  \& Ivanova}{Connelly et~al.}{2012}]{connelly2012absolute}
Connelly J.~N.,  Bizzarro M.,  Krot A.~N.,  Nordlund {\AA}.,  Wielandt D.,   Ivanova M.~A.,  2012, \mn@doi [Science] {10.1126/science.1226919}, 338, 651

\bibitem[\protect\citeauthoryear{Cukanovaite, Tremblay, Toonen, Temmink, Manser, O’Brien  \& McCleery}{Cukanovaite et~al.}{2023}]{cukanovaite2023local}
Cukanovaite E.,  Tremblay P.-E.,  Toonen S.,  Temmink K.,  Manser C.~J.,  O’Brien M.,   McCleery J.,  2023, Monthly Notices of the Royal Astronomical Society, 522, 1643

\bibitem[\protect\citeauthoryear{Da~Silva, Porto~de Mello, Milone, da Silva, Ribeiro  \& Rocha-Pinto}{Da~Silva et~al.}{2012}]{da2012accurate}
Da~Silva R.,  Porto~de Mello G.~F.,  Milone A.,  da Silva L.,  Ribeiro L.,   Rocha-Pinto H.,  2012, \mn@doi [Astronomy \& Astrophysics] {10.1051/0004-6361/201118751}, 542, A84

\bibitem[\protect\citeauthoryear{Dahm}{Dahm}{2015}]{dahm2015reexamining}
Dahm S.,  2015, \mn@doi [The Astrophysical Journal] {10.1088/0004-637X/813/2/108}, 813, 108

\bibitem[\protect\citeauthoryear{Datson, Flynn  \& Portinari}{Datson et~al.}{2015}]{datson2015spectroscopic}
Datson J.,  Flynn C.,   Portinari L.,  2015, \mn@doi [Astronomy \& Astrophysics] {10.1051/0004-6361/201425000}, 574, A124

\bibitem[\protect\citeauthoryear{Del~Peloso, Da~Silva  \& Porto~de Mello}{Del~Peloso et~al.}{2005a}]{del2005agei}
Del~Peloso E.~F.,  Da~Silva L.,   Porto~de Mello G.~F.,  2005a, \mn@doi [Astronomy \& Astrophysics] {10.1051/0004-6361:20047060}, 434, 275

\bibitem[\protect\citeauthoryear{Del~Peloso, Da~Silva, Porto~de Mello  \& Arany-Prado}{Del~Peloso et~al.}{2005b}]{del2005ageiii}
Del~Peloso E.~F.,  Da~Silva L.,  Porto~de Mello G.~F.,   Arany-Prado L.,  2005b, \mn@doi [Astronomy \& Astrophysics] {10.1051/0004-6361:20053307}, 440, 1153

\bibitem[\protect\citeauthoryear{Dopcke, Porto~de Mello  \& Sneden}{Dopcke et~al.}{2019}]{dopcke2019ursa}
Dopcke G.,  Porto~de Mello G.~F.,   Sneden C.,  2019, \mn@doi [Monthly Notices of the Royal Astronomical Society] {10.1093/mnras/stz631}, 485, 4375

\bibitem[\protect\citeauthoryear{Douglas et~al.,}{Douglas et~al.}{2014}]{douglas2014factory}
Douglas S.,  et~al., 2014, The Astrophysical Journal, 795, 161

\bibitem[\protect\citeauthoryear{Duncan et~al.,}{Duncan et~al.}{1991}]{duncan1991ii}
Duncan D.~K.,  et~al., 1991, \mn@doi [Astrophysical Journal Supplement Series] {10.1086/191572}, 76, 383

\bibitem[\protect\citeauthoryear{Dutra-Ferreira, Pasquini, Smiljanic, Porto~de Mello  \& Steffen}{Dutra-Ferreira et~al.}{2016}]{dutra2016consistent}
Dutra-Ferreira L.,  Pasquini L.,  Smiljanic R.,  Porto~de Mello G.~F.,   Steffen M.,  2016, \mn@doi [Astronomy \& Astrophysics] {10.1051/0004-6361/201526783}, 585, A75

\bibitem[\protect\citeauthoryear{Favata, Micela  \& Sciortino}{Favata et~al.}{1997}]{favata1997fe}
Favata F.,  Micela G.,   Sciortino S.,  1997, Astronomy and Astrophysics, 323, 809

\bibitem[\protect\citeauthoryear{Flower}{Flower}{1996}]{flower1996transformations}
Flower P.~J.,  1996, \mn@doi [The Astrophysical Journal] {10.1086/177785}, 469, 355

\bibitem[\protect\citeauthoryear{Fuhrmann, Axer  \& Gehren}{Fuhrmann et~al.}{1993}]{fuhrmann1993balmer}
Fuhrmann K.,  Axer M.,   Gehren T.,  1993, Astronomy and Astrophysics, 271, 451

\bibitem[\protect\citeauthoryear{{Gaia Collaboration} et~al.,}{{Gaia Collaboration} et~al.}{2018}]{2018A&A...616A...1G}
{Gaia Collaboration} et~al., 2018, \mn@doi [\aap] {10.1051/0004-6361/201833051}, \href {https://ui.adsabs.harvard.edu/abs/2018A&A...616A...1G} {616, A1}

\bibitem[\protect\citeauthoryear{Gancarz \& Wasserburg}{Gancarz \& Wasserburg}{1977}]{gancarz1977initial}
Gancarz A.,  Wasserburg G.,  1977, \mn@doi [Geochimica et Cosmochimica Acta] {10.1016/0016-7037(77)90073-4}, 41, 1283

\bibitem[\protect\citeauthoryear{Gehren, Liang, Shi, Zhang  \& Zhao}{Gehren et~al.}{2004}]{gehren2004abundances}
Gehren T.,  Liang Y.,  Shi J.,  Zhang H.,   Zhao G.,  2004, \mn@doi [Astronomy \& Astrophysics] {10.1051/0004-6361:20031582}, 413, 1045

\bibitem[\protect\citeauthoryear{Ghezzi}{Ghezzi}{2005}]{ghezzi2005}
Ghezzi L.,  2005, Monografia de Conclusão de Curso: Projeto SOL (Solar Origin and Life): A Busca do Sol no Tempo, Observatório do Valongo/UFRJ

\bibitem[\protect\citeauthoryear{Gialluca, Robinson, Rugheimer  \& Wunderlich}{Gialluca et~al.}{2021}]{gialluca2021characterizing}
Gialluca M.~T.,  Robinson T.~D.,  Rugheimer S.,   Wunderlich F.,  2021, \mn@doi [Publications of the Astronomical Society of the Pacific] {10.1088/1538-3873/abf367}, 133, 054401

\bibitem[\protect\citeauthoryear{Giribaldi, Ubaldo-Melo, Porto~de Mello, Pasquini, Ludwig, Ulmer-Moll  \& Lorenzo-Oliveira}{Giribaldi et~al.}{2019}]{giribaldi2019accurate}
Giribaldi R.~E.,  Ubaldo-Melo M.~L.,  Porto~de Mello G.~F.,  Pasquini L.,  Ludwig H.-G.,  Ulmer-Moll S.,   Lorenzo-Oliveira D.,  2019, \mn@doi [Astronomy \& Astrophysics] {10.1051/0004-6361/201833763}, 624, A10

\bibitem[\protect\citeauthoryear{Giribaldi, da Silva, Smiljanic  \& Espinoza}{Giribaldi et~al.}{2021}]{giribaldi2021titans}
Giribaldi R.~E.,  da Silva A.~R.,  Smiljanic R.,   Espinoza D.~C.,  2021, Astronomy \& Astrophysics, 650, A194

\bibitem[\protect\citeauthoryear{Gomes~da Silva et~al.,}{Gomes~da Silva et~al.}{2021}]{daSilva2021stellar}
Gomes~da Silva J.,  et~al., 2021, \mn@doi [Astronomy \& Astrophysics] {10.1051/0004-6361/202039765}, 646, A77

\bibitem[\protect\citeauthoryear{Gratton, Carretta, Claudi, Lucatello  \& Barbieri}{Gratton et~al.}{2003}]{gratton2003abundances}
Gratton R.~G.,  Carretta E.,  Claudi R.,  Lucatello S.,   Barbieri M.,  2003, \mn@doi [Astronomy \& Astrophysics] {10.1051/0004-6361:20030439}, 404, 187

\bibitem[\protect\citeauthoryear{Gray, Corbally, Garrison, McFadden  \& Robinson}{Gray et~al.}{2003}]{gray2003contributions}
Gray R.,  Corbally C.,  Garrison R.,  McFadden M.,   Robinson P.,  2003, \mn@doi [The Astronomical Journal] {10.1086/378365}, 126, 2048

\bibitem[\protect\citeauthoryear{Gray, Corbally, Garrison, McFadden, Bubar, McGahee, O’Donoghue  \& Knox}{Gray et~al.}{2006}]{gray2006contributions}
Gray R.,  Corbally C.,  Garrison R.,  McFadden M.,  Bubar E.,  McGahee C.,  O’Donoghue A.,   Knox E.,  2006, \mn@doi [The Astronomical Journal] {10.1086/504637}, 132, 161

\bibitem[\protect\citeauthoryear{Gustafsson, Edvardsson, Eriksson, J{\o}rgensen, Nordlund  \& Plez}{Gustafsson et~al.}{2008}]{gustafsson2008grid}
Gustafsson B.,  Edvardsson B.,  Eriksson K.,  J{\o}rgensen U.~G.,  Nordlund {\AA}.,   Plez B.,  2008, \mn@doi [Astronomy \& Astrophysics] {10.1051/0004-6361:200809724}, 486, 951

\bibitem[\protect\citeauthoryear{Hartmann, Soderblom, Noyes, Burnham  \& Vaughan}{Hartmann et~al.}{1984}]{hartmann1984analysis}
Hartmann L.,  Soderblom D.,  Noyes R.,  Burnham N.,   Vaughan A.,  1984, \mn@doi [The Astrophysical Journal] {10.1086/161609}, 276, 254

\bibitem[\protect\citeauthoryear{Henry, Soderblom, Donahue  \& Baliunas}{Henry et~al.}{1996}]{henry1996survey}
Henry T.~J.,  Soderblom D.~R.,  Donahue R.~A.,   Baliunas S.~L.,  1996, \mn@doi [The Astronomical Journal] {10.1086/117796}, 111

\bibitem[\protect\citeauthoryear{Herbig}{Herbig}{1985}]{herbig1985chromospheric}
Herbig G.,  1985, \mn@doi [The Astrophysical Journal] {10.1086/162887}, 289, 269

\bibitem[\protect\citeauthoryear{Holland}{Holland}{2006}]{holland2006oxygenation}
Holland H.~D.,  2006, \mn@doi [Philosophical Transactions of the Royal Society B: Biological Sciences] {10.1098/rstb.2006.1838}, 361, 903

\bibitem[\protect\citeauthoryear{Holmberg, Nordstr{\"o}m  \& Andersen}{Holmberg et~al.}{2007}]{holmberg2007geneva}
Holmberg J.,  Nordstr{\"o}m B.,   Andersen J.,  2007, \mn@doi [Astronomy \& Astrophysics] {10.1051/0004-6361:20077221}, 475, 519

\bibitem[\protect\citeauthoryear{Huber et~al.,}{Huber et~al.}{2022}]{huber202220}
Huber D.,  et~al., 2022, The Astronomical Journal, 163, 79

\bibitem[\protect\citeauthoryear{Isaacson \& Fischer}{Isaacson \& Fischer}{2010}]{isaacson2010chromospheric}
Isaacson H.,  Fischer D.,  2010, \mn@doi [The Astrophysical Journal] {10.1088/0004-637X/725/1/875}, 725, 875

\bibitem[\protect\citeauthoryear{Jeffries et~al.,}{Jeffries et~al.}{2023}]{jeffries2023gaia}
Jeffries R.,  et~al., 2023, Monthly Notices of the Royal Astronomical Society, 523, 802

\bibitem[\protect\citeauthoryear{Jenkins et~al.,}{Jenkins et~al.}{2006}]{jenkins2006activity}
Jenkins J.,  et~al., 2006, \mn@doi [Monthly Notices of the Royal Astronomical Society] {10.1111/j.1365-2966.2006.10811.x}, 372, 163

\bibitem[\protect\citeauthoryear{Jenkins, Jones, Pavlenko, Pinfield, Barnes  \& Lyubchik}{Jenkins et~al.}{2008}]{jenkins2008metallicities}
Jenkins J.,  Jones H.,  Pavlenko Y.,  Pinfield D.,  Barnes J.,   Lyubchik Y.,  2008, \mn@doi [Astronomy \& Astrophysics] {10.1051/0004-6361:20078611}, 485, 571

\bibitem[\protect\citeauthoryear{Jofr{\'e}, Heiter, Tucci~Maia, Soubiran, Worley, Hawkins, Blanco-Cuaresma  \& Rodrigo}{Jofr{\'e} et~al.}{2018}]{jofre2018gaia}
Jofr{\'e} P.,  Heiter U.,  Tucci~Maia M.,  Soubiran C.,  Worley C.~C.,  Hawkins K.,  Blanco-Cuaresma S.,   Rodrigo C.,  2018, Research Notes of the American Astronomical Society, 2, 152

\bibitem[\protect\citeauthoryear{Joyce \& Chaboyer}{Joyce \& Chaboyer}{2018}]{joyce2018classically}
Joyce M.,  Chaboyer B.,  2018, \mn@doi [The Astrophysical Journal] {10.3847/1538-4357/aad464}, 864, 99

\bibitem[\protect\citeauthoryear{Kounkel et~al.,}{Kounkel et~al.}{2019}]{kounkel2019close}
Kounkel M.,  et~al., 2019, \mn@doi [The Astronomical Journal] {10.3847/1538-3881/ab13b1}, 157, 196

\bibitem[\protect\citeauthoryear{Leenaarts, Carlsson  \& Van Der~Voort}{Leenaarts et~al.}{2012}]{leenaarts2012formation}
Leenaarts J.,  Carlsson M.,   Van Der~Voort L.~R.,  2012, \mn@doi [The Astrophysical Journal] {10.1088/0004-637X/749/2/136}, 749, 136

\bibitem[\protect\citeauthoryear{Linsky, Worden, McClintock  \& Robertson}{Linsky et~al.}{1979}]{linsky1979stellar}
Linsky J.~L.,  Worden S.,  McClintock W.,   Robertson R.~M.,  1979, \mn@doi [Astrophysical Journal Supplement Series, vol. 41, Sept. 1979, p. 47-74.] {10.3847/1538-4357/ab6173}, 41, 47

\bibitem[\protect\citeauthoryear{Lorenzo-Oliveira, Porto~de Mello  \& Schiavon}{Lorenzo-Oliveira et~al.}{2016a}]{lorenzo2016age}
Lorenzo-Oliveira D.,  Porto~de Mello G.~F.,   Schiavon R.~P.,  2016a, \mn@doi [Astronomy \& Astrophysics] {10.1051/0004-6361/201629233}, 594, L3

\bibitem[\protect\citeauthoryear{Lorenzo-Oliveira, Porto~de Mello, Dutra-Ferreira  \& Ribas}{Lorenzo-Oliveira et~al.}{2016b}]{lorenzo2016fine}
Lorenzo-Oliveira D.,  Porto~de Mello G.~F.,  Dutra-Ferreira L.,   Ribas I.,  2016b, \mn@doi [Astronomy \& Astrophysics] {10.1051/0004-6361/201628825}, 595, A11

\bibitem[\protect\citeauthoryear{Lorenzo-Oliveira et~al.,}{Lorenzo-Oliveira et~al.}{2018}]{lorenzo2018solar}
Lorenzo-Oliveira D.,  et~al., 2018, \mn@doi [Astronomy \& Astrophysics] {10.1051/0004-6361/201629294}, 619, A73

\bibitem[\protect\citeauthoryear{Luck}{Luck}{2017}]{luck2017abundances}
Luck R.~E.,  2017, \mn@doi [The Astronomical Journal] {10.3847/1538-3881/153/1/21}, 153, 21

\bibitem[\protect\citeauthoryear{Luck}{Luck}{2018}]{luck2018abundances}
Luck R.~E.,  2018, \mn@doi [The Astronomical Journal] {10.3847/1538-3881/aaa9b5}, 155, 111

\bibitem[\protect\citeauthoryear{Lyra \& Porto~de Mello}{Lyra \& Porto~de Mello}{2005}]{lyra2005fine}
Lyra W.,  Porto~de Mello G.~F.,  2005, \mn@doi [Astronomy \& Astrophysics] {10.1051/0004-6361:20040249}, 431, 329

\bibitem[\protect\citeauthoryear{Mamajek \& Hillenbrand}{Mamajek \& Hillenbrand}{2008}]{mamajek2008improved}
Mamajek E.~E.,  Hillenbrand L.~A.,  2008, \mn@doi [The Astrophysical Journal] {10.1086/591785}, 687, 1264

\bibitem[\protect\citeauthoryear{Mamajek, Meyer  \& Liebert}{Mamajek et~al.}{2002}]{mamajek2002post}
Mamajek E.~E.,  Meyer M.~R.,   Liebert J.,  2002, The Astronomical Journal, 124, 1670

\bibitem[\protect\citeauthoryear{M{\'e}ndez et~al.,}{M{\'e}ndez et~al.}{2021}]{mendez2021habitability}
M{\'e}ndez A.,  et~al., 2021, \mn@doi [Astrobiology] {10.1089/ast.2020.2342}, 21, 1017

\bibitem[\protect\citeauthoryear{Messina et~al.,}{Messina et~al.}{2016}]{messina2016rotation}
Messina S.,  et~al., 2016, \mn@doi [Astronomy \& Astrophysics] {10.1051/0004-6361/201628524}, 596, A29

\bibitem[\protect\citeauthoryear{Metcalfe et~al.,}{Metcalfe et~al.}{2020}]{metcalfe2020evolution}
Metcalfe T.~S.,  et~al., 2020, The Astrophysical Journal, 900, 154

\bibitem[\protect\citeauthoryear{Metcalfe et~al.,}{Metcalfe et~al.}{2023}]{metcalfe2023asteroseismology}
Metcalfe T.~S.,  et~al., 2023, The Astronomical Journal, 166, 167

\bibitem[\protect\citeauthoryear{Metcalfe et~al.,}{Metcalfe et~al.}{2024}]{metcalfe2024weakened}
Metcalfe T.~S.,  et~al., 2024, The Astrophysical Journal Letters, 960, L6

\bibitem[\protect\citeauthoryear{Middelkoop}{Middelkoop}{1982}]{middelkoop1982magnetic}
Middelkoop F.,  1982, Astronomy and Astrophysics, 107, 31

\bibitem[\protect\citeauthoryear{Mishenina, Soubiran, Kovtyukh, Katsova  \& Livshits}{Mishenina et~al.}{2012}]{mishenina2012activity}
Mishenina T.,  Soubiran C.,  Kovtyukh V.,  Katsova M.,   Livshits M.,  2012, \mn@doi [Astronomy \& Astrophysics] {10.1051/0004-6361/201118412}, 547, A106

\bibitem[\protect\citeauthoryear{Mishenina, Pignatari, Korotin, Soubiran, Charbonnel, Thielemann, Gorbaneva  \& Basak}{Mishenina et~al.}{2013}]{mishenina2013abundances}
Mishenina T.,  Pignatari M.,  Korotin S.,  Soubiran C.,  Charbonnel C.,  Thielemann F.-K.,  Gorbaneva T.,   Basak N.~Y.,  2013, \mn@doi [Astronomy \& astrophysics] {10.1051/0004-6361/201220687}, 552, A128

\bibitem[\protect\citeauthoryear{Montes et~al.,}{Montes et~al.}{2018}]{montes2018calibrating}
Montes D.,  et~al., 2018, \mn@doi [Monthly Notices of the Royal Astronomical Society] {10.1093/mnras/sty1295}, 479, 1332

\bibitem[\protect\citeauthoryear{Mosser, Deheuvels, Michel, Th{\'e}venin, Dupret, Samadi, Barban  \& Goupil}{Mosser et~al.}{2008}]{mosser2008hd}
Mosser B.,  Deheuvels S.,  Michel E.,  Th{\'e}venin F.,  Dupret M.-A.,  Samadi R.,  Barban C.,   Goupil M.,  2008, \mn@doi [Astronomy \& Astrophysics] {10.1051/0004-6361:200810011}, 488, 635

\bibitem[\protect\citeauthoryear{Nielsen \& Close}{Nielsen \& Close}{2010}]{nielsen2010uniform}
Nielsen E.~L.,  Close L.~M.,  2010, The Astrophysical Journal, 717, 878

\bibitem[\protect\citeauthoryear{Noyes, Hartmann, Baliunas, Duncan  \& Vaughan}{Noyes et~al.}{1984}]{noyes1984rotation}
Noyes R.,  Hartmann L.,  Baliunas S.,  Duncan D.,   Vaughan A.,  1984, \mn@doi [The Astrophysical Journal] {10.1086/161945}, 279, 763

\bibitem[\protect\citeauthoryear{Olsen}{Olsen}{1983}]{olsen1983four}
Olsen E.,  1983, Astronomy and Astrophysics Supplement Series, 54, 55

\bibitem[\protect\citeauthoryear{Olsen}{Olsen}{1993}]{olsen1993stromgren}
Olsen E.,  1993, Astronomy and Astrophysics Supplement Series, 102, 89

\bibitem[\protect\citeauthoryear{Olsen}{Olsen}{1994}]{olsen1994large}
Olsen E.,  1994, Astronomy and Astrophysics Supplement Series, 104, 429

\bibitem[\protect\citeauthoryear{Pace}{Pace}{2013}]{pace2013chromospheric}
Pace G.,  2013, \mn@doi [Astronomy \& Astrophysics] {10.1051/0004-6361/201220364}, 551, L8

\bibitem[\protect\citeauthoryear{Pace \& Pasquini}{Pace \& Pasquini}{2004}]{pace2004age}
Pace G.,  Pasquini L.,  2004, \mn@doi [Astronomy \& Astrophysics] {10.1051/0004-6361:20040568}, 426, 1021

\bibitem[\protect\citeauthoryear{Pasquini \& Pallavicini}{Pasquini \& Pallavicini}{1991}]{pasquini1991h}
Pasquini L.,  Pallavicini R.,  1991, Astronomy and Astrophysics, 251, 199

\bibitem[\protect\citeauthoryear{Pasquini, Pala, Salaris, Ludwig, Leao, Weiss  \& de Medeiros}{Pasquini et~al.}{2023}]{pasquini2023accurate}
Pasquini L.,  Pala A.,  Salaris M.,  Ludwig H.,  Leao I.,  Weiss A.,   de Medeiros J.,  2023, Monthly Notices of the Royal Astronomical Society, 522, 3710

\bibitem[\protect\citeauthoryear{Paulson \& Yelda}{Paulson \& Yelda}{2006}]{paulson2006differential}
Paulson D.~B.,  Yelda S.,  2006, \mn@doi [Publications of the Astronomical Society of the Pacific] {10.1086/504115}, 118, 706

\bibitem[\protect\citeauthoryear{Paulson, Sneden  \& Cochran}{Paulson et~al.}{2003}]{paulson2003searching}
Paulson D.~B.,  Sneden C.,   Cochran W.~D.,  2003, \mn@doi [The Astronomical Journal] {10.1086/375209}, 125, 3185

\bibitem[\protect\citeauthoryear{Porto~de Mello, Lyra  \& Keller}{Porto~de Mello et~al.}{2008}]{de2008alpha}
Porto~de Mello G.,  Lyra W.,   Keller G.,  2008, \mn@doi [Astronomy \& Astrophysics] {10.1051/0004-6361:200810031}, 488, 653

\bibitem[\protect\citeauthoryear{Porto~de Mello, Da~Silva, Da~Silva  \& De~Nader}{Porto~de Mello et~al.}{2014}]{de2014photometric}
Porto~de Mello G.,  Da~Silva R.,  Da~Silva L.,   De~Nader R.,  2014, \mn@doi [Astronomy \& Astrophysics] {10.1051/0004-6361/201322277}, 563, A52

\bibitem[\protect\citeauthoryear{Ram{\'\i}rez, Fish, Lambert  \& Prieto}{Ram{\'\i}rez et~al.}{2012}]{ramirez2012lithium}
Ram{\'\i}rez I.,  Fish J.,  Lambert D.~L.,   Prieto C.~A.,  2012, \mn@doi [The Astrophysical Journal] {10.1088/0004-637X/756/1/46}, 756, 46

\bibitem[\protect\citeauthoryear{Ram{\'\i}rez, Prieto  \& Lambert}{Ram{\'\i}rez et~al.}{2013}]{ramirez2013oxygen}
Ram{\'\i}rez I.,  Prieto C.~A.,   Lambert D.~L.,  2013, \mn@doi [The Astrophysical Journal] {10.1088/0004-637X/764/1/78}, 764, 78

\bibitem[\protect\citeauthoryear{Randich, Gratton, Pallavicini, Pasquini  \& Carretta}{Randich et~al.}{1999}]{randich1999lithium}
Randich S.,  Gratton R.,  Pallavicini R.,  Pasquini L.,   Carretta E.,  1999, Astronomy and Astrophysics, 348, 487

\bibitem[\protect\citeauthoryear{Rocha-Pinto \& Maciel}{Rocha-Pinto \& Maciel}{1998}]{rocha1998metallicity}
Rocha-Pinto H.,  Maciel W.~J.,  1998, \mn@doi [Monthly Notices of the Royal Astronomical Society] {10.1046/j.1365-8711.1998.01597.x}, 298, 332

\bibitem[\protect\citeauthoryear{Santos, Israelian  \& Mayor}{Santos et~al.}{2004}]{santos2004spectroscopic}
Santos N.~C.,  Israelian G.,   Mayor M.,  2004, \mn@doi [Astronomy \& Astrophysics] {10.1051/0004-6361:20034469}, 415, 1153

\bibitem[\protect\citeauthoryear{Santos, Israelian, Mayor, Bento, Almeida, Sousa  \& Ecuvillon}{Santos et~al.}{2005}]{santos2005spectroscopic}
Santos N.~C.,  Israelian G.,  Mayor M.,  Bento J.,  Almeida P.,  Sousa S.,   Ecuvillon A.,  2005, \mn@doi [Astronomy \& Astrophysics] {10.1051/0004-6361:20052895}, 437, 1127

\bibitem[\protect\citeauthoryear{Schoolman}{Schoolman}{1972}]{schoolman1972formation}
Schoolman S.~A.,  1972, \mn@doi [Solar Physics] {10.1007/BF00148701}, 22, 344

\bibitem[\protect\citeauthoryear{Schrijver}{Schrijver}{1995}]{schrijver1995basal}
Schrijver C.~J.,  1995, \mn@doi [The Astronomy and Astrophysics Review] {10.1007/BF01837115}, 6, 181

\bibitem[\protect\citeauthoryear{Schrijver}{Schrijver}{2023}]{schrijver2023stellar}
Schrijver C.,  2023, \mn@doi [Solar Physics] {10.1007/s11207-023-02200-y}, 298, 104

\bibitem[\protect\citeauthoryear{Schr{\"o}der, Reiners  \& Schmitt}{Schr{\"o}der et~al.}{2009}]{schroder2009ii}
Schr{\"o}der C.,  Reiners A.,   Schmitt J.,  2009, \mn@doi [Astronomy \& Astrophysics] {10.1051/0004-6361:200810377}, 493, 1099

\bibitem[\protect\citeauthoryear{Schuler, Plunkett, King  \& Pinsonneault}{Schuler et~al.}{2010}]{schuler2010fe}
Schuler S.~C.,  Plunkett A.~L.,  King J.~R.,   Pinsonneault M.~H.,  2010, \mn@doi [Publications of the Astronomical Society of the Pacific] {10.1086/655026}, 122, 766

\bibitem[\protect\citeauthoryear{Sissa et~al.,}{Sissa et~al.}{2016}]{sissa2016halpha}
Sissa E.,  et~al., 2016, \mn@doi [Astronomy \& Astrophysics] {10.1051/0004-6361/201628531}, 596, A76

\bibitem[\protect\citeauthoryear{{Sitnova} et~al.,}{{Sitnova} et~al.}{2015}]{sitnova2015systematic}
{Sitnova} T.,  et~al., 2015, \mn@doi [The Astrophysical Journal] {10.1088/0004-637X/808/2/148}, 808, 148

\bibitem[\protect\citeauthoryear{Skumanich}{Skumanich}{1972}]{skumanich1972time}
Skumanich A.,  1972, \mn@doi [The Astrophysical Journal] {10.1086/151310}, 171, 565

\bibitem[\protect\citeauthoryear{Soderblom}{Soderblom}{1985}]{soderblom1985survey}
Soderblom D.~R.,  1985, \mn@doi [The Astronomical Journal] {10.1086/113918}, 90, 2103

\bibitem[\protect\citeauthoryear{Soderblom}{Soderblom}{2010}]{soderblom2010ages}
Soderblom D.~R.,  2010, Annual Review of Astronomy and Astrophysics, 48, 581

\bibitem[\protect\citeauthoryear{Soderblom, Duncan  \& Johnson}{Soderblom et~al.}{1991}]{soderblom1991chromospheric}
Soderblom D.~R.,  Duncan D.~K.,   Johnson D.~R.,  1991, \mn@doi [The Astrophysical Journal] {10.1086/170238}, 375, 722

\bibitem[\protect\citeauthoryear{Soriano \& Vauclair}{Soriano \& Vauclair}{2010}]{soriano2010new}
Soriano M.,  Vauclair S.,  2010, \mn@doi [Astronomy \& Astrophysics] {10.1051/0004-6361/200911862}, 513, A49

\bibitem[\protect\citeauthoryear{Soto \& Jenkins}{Soto \& Jenkins}{2018}]{soto2018spectroscopic}
Soto M.,  Jenkins J.~S.,  2018, \mn@doi [Astronomy \& Astrophysics] {10.1051/0004-6361/201731533}, 615, A76

\bibitem[\protect\citeauthoryear{Sousa, Santos, Israelian, Mayor  \& Monteiro}{Sousa et~al.}{2006}]{sousa2006spectroscopic}
Sousa S.,  Santos N.,  Israelian G.,  Mayor M.,   Monteiro M.,  2006, \mn@doi [Astronomy \& Astrophysics] {10.1051/0004-6361:20065658}, 458, 873

\bibitem[\protect\citeauthoryear{Stanford-Moore, Nielsen, De~Rosa, Macintosh  \& Czekala}{Stanford-Moore et~al.}{2020}]{stanford2020baffles}
Stanford-Moore S.~A.,  Nielsen E.~L.,  De~Rosa R.~J.,  Macintosh B.,   Czekala I.,  2020, The Astrophysical Journal, 898, 27

\bibitem[\protect\citeauthoryear{Steenbock}{Steenbock}{1983}]{steenbock1983barium}
Steenbock W.,  1983, Astronomy and Astrophysics, 126, 325

\bibitem[\protect\citeauthoryear{Tagliaferri, Cutispoto, Pallavicini, Randich  \& Pasquini}{Tagliaferri et~al.}{1994}]{tagliaferri1994photometric}
Tagliaferri G.,  Cutispoto G.,  Pallavicini R.,  Randich S.,   Pasquini L.,  1994, Astronomy \& Astrophysics, 285, 272

\bibitem[\protect\citeauthoryear{Tang \& Gai}{Tang \& Gai}{2011}]{tang2011asteroseismic}
Tang Y.,  Gai N.,  2011, \mn@doi [Astronomy \& Astrophysics] {10.1051/0004-6361/201014886}, 526, A35

\bibitem[\protect\citeauthoryear{Torres}{Torres}{2010}]{torres2010use}
Torres G.,  2010, \mn@doi [The Astronomical Journal] {10.1088/0004-6256/140/5/1158}, 140, 1158

\bibitem[\protect\citeauthoryear{Trevisan, Barbuy, Eriksson, Gustafsson, Grenon  \& Pomp{\'e}ia}{Trevisan et~al.}{2011}]{trevisan2011analysis}
Trevisan M.,  Barbuy B.,  Eriksson K.,  Gustafsson B.,  Grenon M.,   Pomp{\'e}ia L.,  2011, \mn@doi [Astronomy \& Astrophysics] {10.1051/0004-6361/201016056}, 535, A42

\bibitem[\protect\citeauthoryear{Valenti \& Fischer}{Valenti \& Fischer}{2005}]{valenti2005spectroscopic}
Valenti J.~A.,  Fischer D.~A.,  2005, \mn@doi [The Astrophysical Journal Supplement Series] {10.1086/430500}, 159, 141

\bibitem[\protect\citeauthoryear{Van~Leeuwen}{Van~Leeuwen}{2007}]{van2007validation}
Van~Leeuwen F.,  2007, \mn@doi [Astronomy \& Astrophysics] {10.1051/0004-6361:20078357}, 474, 653

\bibitem[\protect\citeauthoryear{Vernazza, Avrett  \& Loeser}{Vernazza et~al.}{1981}]{vernazza1981structure}
Vernazza J.~E.,  Avrett E.~H.,   Loeser R.,  1981, \mn@doi [Astrophysical Journal Supplement Series, vol. 45, Apr. 1981, p. 635-725.] {10.1086/190731}, 45, 635

\bibitem[\protect\citeauthoryear{Wittenmyer, Liu, Wang, Casagrande, Johnson  \& Tinney}{Wittenmyer et~al.}{2016}]{wittenmyer2016pan}
Wittenmyer R.~A.,  Liu F.,  Wang L.,  Casagrande L.,  Johnson J.~A.,   Tinney C.,  2016, \mn@doi [The Astronomical Journal] {10.3847/0004-6256/152/1/19}, 152, 19

\bibitem[\protect\citeauthoryear{Wright, Marcy, Butler  \& Vogt}{Wright et~al.}{2004a}]{wright2004chromospheric}
Wright J.~T.,  Marcy G.~W.,  Butler R.~P.,   Vogt S.~S.,  2004a, \mn@doi [The Astrophysical Journal Supplement Series] {10.1086/386283}, 152, 261

\bibitem[\protect\citeauthoryear{Wright, Marcy, Butler  \& Vogt}{Wright et~al.}{2004b}]{wright2004}
Wright J.~T.,  Marcy G.~W.,  Butler R.~P.,   Vogt S.~S.,  2004b, The Astrophysical Journal Supplement Series, 152, 261

\bibitem[\protect\citeauthoryear{Zhang, Zhao, Oswalt, Fang, Zhao, Liang, Ye  \& Zhong}{Zhang et~al.}{2019}]{zhang2019stellar}
Zhang J.,  Zhao J.,  Oswalt T.~D.,  Fang X.,  Zhao G.,  Liang X.,  Ye X.,   Zhong J.,  2019, \mn@doi [The Astrophysical Journal] {10.3847/1538-4357/ab4efe}, 887, 84

\bibitem[\protect\citeauthoryear{Zhao \& Magain}{Zhao \& Magain}{1991}]{zhao1991abundances}
Zhao G.,  Magain P.,  1991, Astronomy and Astrophysics, 244, 425

\makeatother
\end{thebibliography}








\bsp	
\label{lastpage}
\end{document}